\documentclass[twocolumn,showpacs,preprintnumbers,amsmath,amssymb]{revtex4}

\usepackage{CJK}

\usepackage[unicode]{hyperref}
\usepackage[caption=false]{subfig}
\usepackage{graphicx}
\usepackage{color} 
\graphicspath{ {./images/} }
\usepackage{dcolumn}
\usepackage{bm}
\usepackage[caption=false]{subfig}
\newcommand{\half}{\mbox{$\frac12$}}
\let\hat=\widehat
\let\tilde=\widetilde
\newcommand{\dd}{\mbox{d}}

\newcommand{\qq}[1]{#1}

\renewcommand{\theequation}{\arabic{section}.\arabic{equation}}

\let\hat=\widehat
\let\tilde=\widetilde 
\newfont{\bbold}{msbm10 scaled\magstep1}

\newcommand{\nn}{\nonumber}
\newcommand{\ro}{\rho\,}

\newcommand{\ep}{\epsilon}

\newcommand{\rec}[1]{\mbox{$\frac{1}{#1}$}}
\newcommand{\mfrac}[2]{\mbox{$\frac{#1}{#2}$}}
\newcommand{\ee}{{\rm e}}

\newcommand{\beq}{\begin{equation}} 
\newcommand{\eeq}{\end{equation}} 
\newcommand{\beqa}{\begin{eqnarray}}
\newcommand{\eeqa}{\end{eqnarray}}

\newcommand{\sech}{{\rm sech}}

\begin{document}

\begin{CJK*}{GB}{}

\preprint{APS/123-QED}

\title{{\bf Breather modes of fully nonlinear mass-in-mass chains}}

\author{Jonathan A.D.~Wattis$^1$ }
\email[$^1$]{ Jonathan.Wattis@nottingham.ac.uk}
\date{\today}

\affiliation{${}^{1}$School of Mathematical Sciences, University 
of Nottingham, \\ University Park, Nottingham NG7 2RD, UK.}

\begin{abstract}
We propose a model for a chain of particles coupled by 
nonlinear springs in which each mass has an internal mass 
and all interactions are assumed to be nonlinear.  We show 
how to construct an asymptotic solution of this system 
using multiple timescales, the systematic solution of coupled 
equations by repeated application of a consistency condition. 
Our results show that for some combinations of nonlinearity 
the dynamics are governed by the NLS as in the more usual 
mass-in-mass chains with linear interactions between 
inner and outer masses.  However, when both 
nonlinearities have quadratic components, we show that 
the asymptotic reduction results in a Ginzburg-Landau 
equation instead of NLS.  
\end{abstract}
 

\maketitle

\end{CJK*} 

\section{Introduction}
\setcounter{equation}{0}

The dynamics exhibited by chains of particles coupled by 
nonlinear springs has been of long-term interest since the 
pioneering study of Fermi, Pasta, Ulam and Tsingou (FPUT) \cite{fpu}.  
Initially, travelling waves were the main focus of interest 
\cite{zk,fw}; however, for the last couple of decades, the 
behaviour of breather-modes in these systems has been 
a key component \cite{ma,chong}, and more recently, the 
types of chain which exihibit these mode has been extended, 
to diatomic chains \cite{cretegny,livi,jw-diFPU}, 
two-dimensional lattices \cite{flach,jce+marin,jce+marin2,
dario2,bajars,butt,butt2,alz}, and mass-in-mass chains. 
For a recent review of the applications of these systems, 
see  Archilla {\em et al.}\ \cite{QinMica}. 

In mass-in-mass systems, the interconnected nodes are 
assumed to contain an internal oscillator, which allows a 
more complicated frequency response. Most commonly, 
the along-chain interactions are assumed to be nonlinear, 
whilst the interactions between inner and outer particles 
are linear as in \cite{dario3,ksx,vain}; however, in 
some cases, the along chain interactions are linear, and 
the inner-outer interactions are nonlinear, for example, 
see Wallen {\em et al}.\ \cite{wallen}.  
Liu {\em et al.}\ \cite{vain} investigate the lifetimes of bright 
breathers in the problem with Hertzian contact by reducing 
the equations to a discrete $p$-Schrodinger equation. 
\qq{Liu {\em et al.}\ \cite{liu} use Schrodinger reductions 
to investigate the form and stability of localised energy 
transport in these systems, they note the existence 
of both bright and dark breathers in alternating regions 
of parameter space.}\ 
Conditions for the existence of travelling waves have been 
explored by Kevrekidis {\em et al.}\ \cite{ksx}. In \cite{dario3}, 
Kevrekidis {\em et al}.\ analyse energy trapping due to a 
localised defect in a Hertzian chain with internal masses. 
\qq{Bonanomi {\em et al.}\ \cite{bona} also analyse wave 
propagation in chains with internal resonators; they 
observe a wide gap between the frequency bands
corresponding to linear waves.   
The simpler case of a {\em single} resonant defect is 
considered by Lydon {\em et al} \cite{lydon}. }

In this paper \qq{we consider the case where there is an 
internal resonator at every node along the lattice, and}\ 
further generalise these mass-in-mass 
systems to allow both interactions to be nonlinear, that is, 
both between the internal oscillator and external shell, and 
the interaction between neighbouring particles along the chain. 
\qq{One application of such a model is a precompressed 
Hertzian chain, of particles in contact, in which each particle 
contains an identical nonlinear resonator.  Such systems 
clearly have nonlinear nearest-neighbour interactions, which 
can be adjusted by varying the amount of precompression 
applied.  Whilst we acknowledge that Hertzian contact 
may be more strongly nonlinear than an internal resonator, 
no experimental oscillator can be precisely linear, 
so it seems natural to model both internal and 
nearest-neighbour interactions as nonlinear. 
The results we derive below suggest that the effects 
of combining these two nonlinearities can be significant. }\ 
From a mathematical modelling perspective,  \qq{the 
inclusion of nonlinear terms in both interaction forces}\ is a natural 
generalisation by which the mass-in-mass model is extended. 
Much of the previous theoretical analysis of mass-in-mass systems 
has relied on this inner-outer relationship being linear, 
which leads to some simplification of the theory.   
In the analysis presented below, we include nonlinear terms, 
showing how the nonlinear terms can be accommodated in 
a full asymptotic solution of the dynamics using multiple 
scales techniques \cite{bo}.  
We find conditions on the form of the nonlinearities 
required for breathers to be long-lived.

\section{Fully nonlinear mass in mass system}
\setcounter{equation}{0}

\begin{figure} 
\includegraphics[scale=1.0]{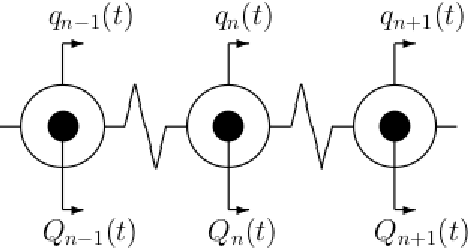}
\caption{Illustration of the mass-in-mass chain 
whose dynamics are described in this paper. 
\label{chain-fig}}
\end{figure}

Figure \ref{chain-fig} illustrates the chain of coupled 
mass-in-mass oscillators that we are modelling.  We define 
the displacements of the outer oscillators of mass $m$ by 
$q_n(t)$, with corresponding momenta $p_n(t)$. 
These are coupled to their nearest neighbours ($n\pm1$), 
as well as the inner masses ($M$), whose displacements 
and momenta are given by $Q_n(t)$, $P_n(t)$. 
We derive the equations of motion from the Hamiltonian 
\beq
H = \sum_n \frac{ p_n^2 }{2m} + \frac{P_n^2}{2M} 
+ V(q_{n+1}-q_n) + W(q_n-Q_n) , 
\eeq
where the potential energies are given by 
\begin{align}
V(\phi) = &\; \half \phi^2 + \rec{3} a \phi^3 
+ \rec{4} b \phi^4, \label{V}\\
W(\psi) = &\; \half \rho \psi^2 + \rec{3} \alpha \psi^3 
+ \rec{4} \beta \psi^4 ,  \label{W}
\end{align}
for some $\rho>0$, with $a,b,\alpha,\beta$ of either sign. 

The equations of motion are then 
\begin{align}
m \frac{\dd^2 q_n}{\dd t^2} =&\; V'(q_{n+1}\!-\!q_n) 
- V'(q_n\!-\!q_{n-1}) - W'(q_n\!-\!Q_n) , 
\nn \\ & \label{q-eq} \\ 
M \frac{\dd^2 Q_n}{\dd t^2} =&\; W'(q_n-Q_n) , 
\label{QQ-eq} \end{align}
where
\begin{align}
V'(\phi) = &\; \phi + a \phi^2 + b \phi^3, \label{Vp}\\
W'(\psi) = &\; \rho\psi + \alpha \psi^2 + \beta \psi^3 , 
\label{Wp} \end{align}
represent the forces due to nearest-neighbour, 
and inner-outer interactions. 
We propose to investigate the form small amplitude 
breathers in this system, using multiple-scales 
asymptotic methods \cite{bo}. 

\section{Asymptotic analysis}
\setcounter{equation}{0}

We seek waves which have the form of a linear wave whose 
amplitude is modulated by a slowly-varying envelope. We introduce 
a small parameter, $\ep$, which is proportional to the amplitude 
of the breather soluton; since we use a multiple scales techniques, 
we introduce a large space scale ($y$) and two long timescales, 
\begin{align}
\ep \ll 1 , \quad y = \ep n , \quad 
\tau = \ep t , \quad T = \ep^2 t . \label{scales}
\end{align}
The leading order linear wave has  the form $q,Q \propto 
\mbox{Re}(\ee^{i\theta}) = \mbox{Re}(\ee^{i(kn-\omega(k)t})$. 
Since we wish to consider quadratic nonlinearities, the centre 
of the oscillation may be offset from zero, so we 
include a `zero'-mode in addition to the envelope that 
describes the amplitude of the oscillations. 
We use $F_j(y,\tau,T),P_j(y,\tau,T)$ for the leading order expressions for 
the amplitude envelope and zero mode, hence our ansatz is 
\begin{align}
q_n(t) =&\; \ep \ee^{i\theta} F_1(y,\tau,T) + \ep F_0(y,\tau,T) \nn \\ 
& \; + \ep^2 \left[ \ee^{2i\theta} G_2 + \ee^{i\theta} G_1 
+ G_0 \right] \nn \\ & 
+ \ep^3 \left[ \ee^{3i\theta} H_3 + 
\ee^{2i\theta} H_2 + \ee^{i\theta} H_1 + H_0 \right] 
\nn \\ & + \ep^4 \left[ I_0 + \ldots \right] 
+ \ldots + c.c. \label{q-ans} , \\ 
Q_n(t) =&\; \ep \ee^{i\theta} P_1(y,\tau,T) + \ep P_0(y,\tau,T) \nn \\ 
&\; + \ep^2 \left[ \ee^{2i\theta} S_2 + \ee^{i\theta} S_1 
+ S_0 \right] \nn \\ & 
+ \ep^3 \left[ \ee^{3i\theta} R_3 + 
\ee^{2i\theta} R_2 + \ee^{i\theta} R_1 + R_0 \right] 
\nn \\ & + \ep^4 \left[ U_0 + \ldots \right] 
+ \ldots + c.c., \label{QQ-ans} 
\end{align}
where $G_j,H_j,S_j,R_j$ describe the amplitudes of other modes 
caused by nonlinearities, which are also functions of $(y,\tau,T)$ 
and are determined by correction terms of higher order in $\ep$. 
These expressions are substituted into the equations of motion 
(\ref{q-eq})--(\ref{QQ-eq}), then all terms are expanded in powers 
of $\ep$.  From (\ref{scales}), the time derivative is expanded as 
$\dd/\dd t = \partial_t + \ep \partial_\tau + \ep^2 \partial_T$.  
Equating terms of equal powers of $\ep$ and equal 
frequencies (in terms of $\ee^{im\theta}$, for $m = 
0,1,2,\ldots$), gives a hierarchy of coupled pairs of equations 
which determine the shape of the envelopes $F_j,P_j,G_j,S_j$, 
etc.; in the remainder of this section we work through the 
systems of equations sequentially. 

\subsection{Equations at $\mathcal{O}(\ep\ee^{i\theta})$
\label{ep1e1-sec}}

\begin{figure}[ht] 
\hspace*{-10mm}
\includegraphics[scale=0.32]{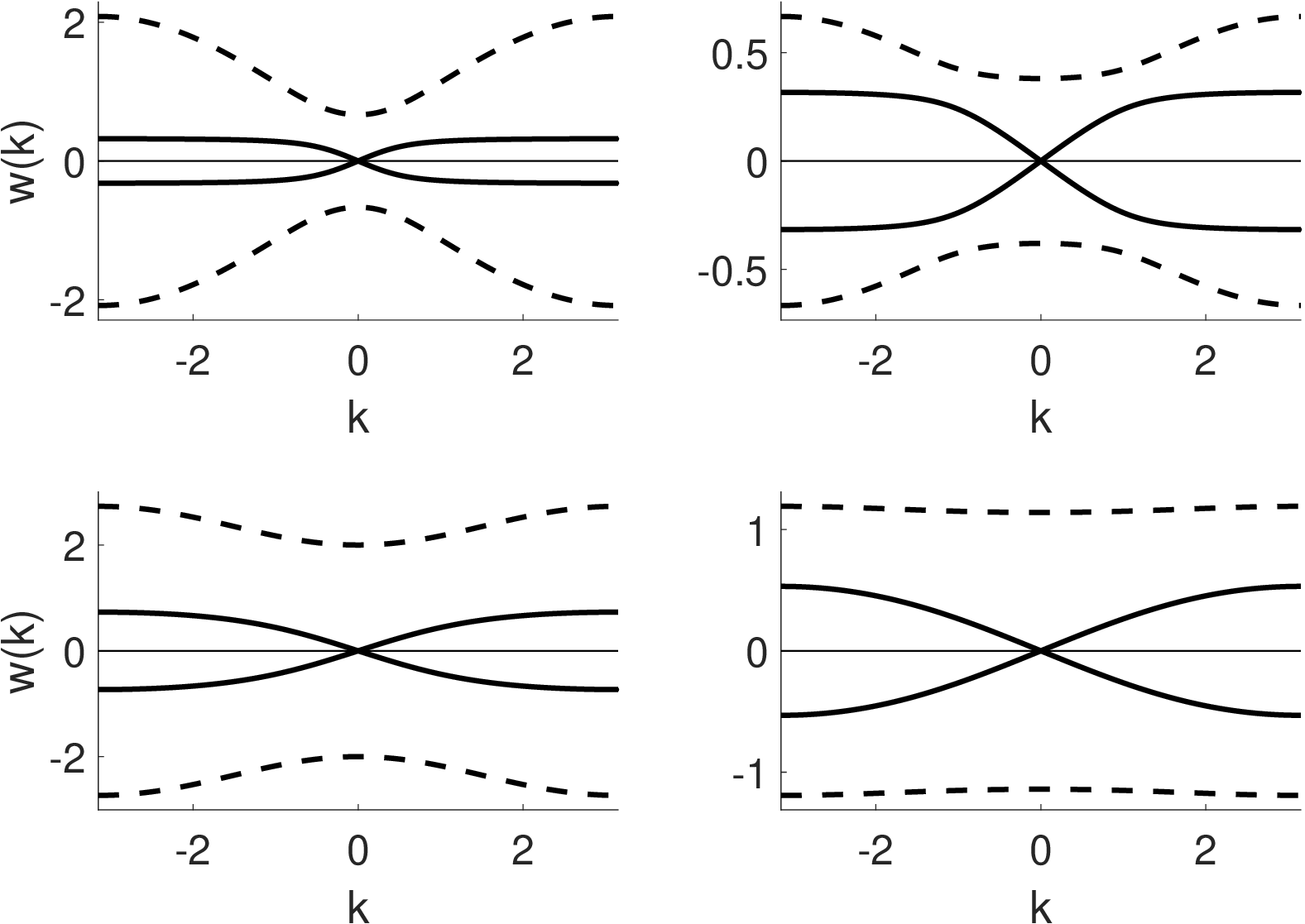} 
\caption{Illustration of the dispersion relation 
(\ref{disp}) for the cases: 
top left: $\rho=1/3$, $\mu=3$ ($m=1$, $M=3$); 
top right: $\rho=1/3$, $\mu=0.3$, ($m=10$, $M=3$); 
bottom left: $\rho=3$, $\mu=3$ ($m=1$, $M=3$); 
bottom right: $\rho=3$, $\mu=0.3$ ($m=10$, $M=3$). 
The solid lines correspond to the acoustic 
mode and the dashed lines to the optical. 
These cases and key apply to the illustrations in all later figures. 
\label{disp-fig} }
\end{figure}

Substituting the ansatz (\ref{q-ans})--(\ref{QQ-ans}) 
into the governing equations (\ref{q-eq})--(\ref{QQ-eq}) 
and expanding, we find, at leading order 
\begin{align} 
{\bf M} \begin{pmatrix} F_1 \\ P_1 \end{pmatrix} := & 
\begin{pmatrix} 
m \omega^2 \!-\! 4 \sin^2\half k \!-\! \rho & \rho 
\\ \rho & M \omega^2 \!-\! \rho \end{pmatrix}
\begin{pmatrix} F_1 \\ P_1 \end{pmatrix} = 
\begin{pmatrix} 0 \\ 0 \end{pmatrix} . 
\nn \\ \label{eq11} \end{align}
For there to be nonzero solutions for $F_1,P_1$, we 
need the matrix to be singular, which occurs when 
the frequency respons, $\omega$, is given by 
\begin{align}
\omega^2 = &\; \frac{1}{2mM} \left[ \ro M \!+\! \ro m 
\!+\! 4M\sin^2 \half k \pm \sqrt{ D } \right] , \nn \\ 
D = &\; 
(\ro M \!+\! \ro m \!+\! 4M\sin^2 \half k)^2 
- 16Mm\ro\sin^2\half k . \nn \\ & 
\label{disp} \end{align}
This relationship is illustrated in Figure \ref{disp-fig} 
for a variety of values of $\ro,m,M$.  Note that there are 
two modes for each frequency (discounting the $\omega 
\mapsto -\omega$ symmetry).  We refer to the one with 
the larger frequency as the optical mode 
($\omega_{\mbox{\scriptsize op}}$), and the smaller 
frequency one as the acoustic mode 
($\omega_{\mbox{\scriptsize ac}}$).  

To give simple explicit examples, we consider the 
asymptotic cases of large mass ratios, defining 
\beq  \mu = M/m ,  \label{mu-def} \eeq
as the ratio of the inner mass to the outer. 
The speed of sound in the lattice is defined by 
$c_0 = \lim_{k\rightarrow0} \omega_{\mbox{\scriptsize ac}}(k)/k$, 
which gives 
\beq
c_0 = \frac{1}{\sqrt{m(1+\mu)}} . \label{co-def} 
\eeq
This speed is small when the mass ratio ($\mu$) is large. 

To illustrate the types of behaviour that may be observed, 
we consider mass ratios (\ref{mu-def}), either side of unity, 
namely 3 and 0.3; and spring constants above and below unity, 
i.e.\ $\ro=1/3, 3$. We also consider cases with no quadratic 
nonlinearities ($\alpha=a=0$) and with both ($\alpha \neq 0 \neq a$), 
as well as with one but not the other (both $\alpha = 0 \neq a$ 
and $\alpha \neq 0 = a$).  

We observe that in many cases there are a large range of 
wavenumbers, $k$, which give rise to almost the same 
frequency ($\omega$). For example, in the lower right panel, 
the optical frequency is almost independent of wavenumber, 
whilst the acoustic mode has a strong dependence on $k$;  
\qq{that is, as the wavenumber $k$ varies from zero to $\pi/2$, 
the acoustic band covers a considerable range of frequencies, 
($0 \leq \omega \leq 0.53$), whereas, in the same range of $k$, 
only a very small range of frequencies are covered 
by the optical band, namely ($1.14 \leq \omega \leq 1.19$ -- 
less than one tenth of range of the acoustic band).

This is in contrast with}\ the top left panel, where the situation 
is reversed: the acoustic mode is almost independent on $k$ 
whilst the optical mode varies significantly with $k$; 
\qq{ here, as $k$ ranges from zero to $\pi/2$,
the acoustic band spans $0 \leq \omega \leq 0.32$ 
whilst the optical band spans $0.67 \leq \omega \leq 2.08$ 
-- about four times the range of the acoustic band. 
In the lower left panel, both modes vary with $k$. 
the acoustic and optical modes spanning 
$0 \leq \omega \leq 0.73$ and $2.00 \leq \omega \leq 2.73$ 
respectively.  Similarly, the top right panel 
also has relatively wide ranges, namely 
$0 \leq \omega \leq 0.32$ and $0.38 \leq \omega \leq 0.67$. 
The size of the gap between the bands is also 
affected strongly by $\ro$ and $\mu$, 
in the four panels the gaps are 0.33,  0.06, 1.27, 0.61 --which we 
note are small in the first two cases--when $\rho$ is small 
and significantly larger when $\rho$ is larger. } 

From Figure \ref{disp-fig} we also note that there is a gap 
between the acoustic and optical modes, and this gap can 
be relatively wide (as in the bottom panels), but also may 
be very small (top right panel). It is the regions above the 
optical mode, and between the acoustic and optical modes 
that we expect breathers to exist and be stable. 
We note that the gap between the two branches is always 
positive, and is given by 
\begin{align}
\Delta \omega = & \; \omega_{\mbox{\scriptsize op}}(0)
- \omega_{\mbox{\scriptsize ac}}(\pi) 
= \frac{\omega_{\mbox{\scriptsize op}}^2(0)
- \omega_{\mbox{\scriptsize ac}}^2(\pi)}{
\omega_{\mbox{\scriptsize op}}(0)
+ \omega_{\mbox{\scriptsize ac}}(\pi)} , 
\end{align}
which is always positive, since the numerator is given by 
\begin{align}
\Delta^2 = & \omega_{\mbox{\scriptsize op}}^2(0)
- \omega_{\mbox{\scriptsize ac}}^2(\pi) \nn \\ 
= & \frac{\ro}{2\mu m} \left[ 1 + \mu - \frac{4\mu}{\ro} + \sqrt{  \left( 
1 + \mu - \frac{4\mu}{\ro} \right)^2 + \frac{16\mu^2}{\ro} }  \right] . 
\end{align} 
For any given mass ratio, $\mu$, the narrowest gap is obtained 
when the spring constant is given by $\ro = 4\mu / (1+\mu)$. 

In the limit of small $\mu$, we find 
\begin{align}
\omega_{\mbox{\scriptsize ac}}^2 \sim &\; 
\frac{4}{m} \sin^2 (\half k) (1-\mu), \quad & 
\omega_{\mbox{\scriptsize op}}^2 \sim &\; 
\frac{\ro (1+\mu) }{ \mu m } , 
\end{align}
whilst for large mass ratio ($\mu\gg1$) we have 
\begin{align}
\omega_{\mbox{\scriptsize ac}}^2 \sim &\; 
\frac{4\ro\sin^2(\half k) }{\mu m (\ro +4 \sin^2(\half k))},& 
\omega_{\mbox{\scriptsize op}}^2 \sim &\; 
\frac{\ro + 4 \sin^2 (\half k)}{m} . \nn \\ &&& 
\end{align}
The small $\mu$ case corresponds to the inner 
oscillators having negligible mass, whereas in case 
of large $\mu$, the inner masses dominate.  Whilst 
we might expect the former case to be a regular 
perturbation of the FPUT system and the latter case 
give rise to more exotic dynamics, the observed 
behaviour will also depend on the strength of the 
interaction between the inner and outer masses, 
$\ro$, and, at larger amplitudes, also $\alpha,\beta$. 

\begin{figure}[!ht] 
\hspace*{-5mm}
\includegraphics[scale=0.31]{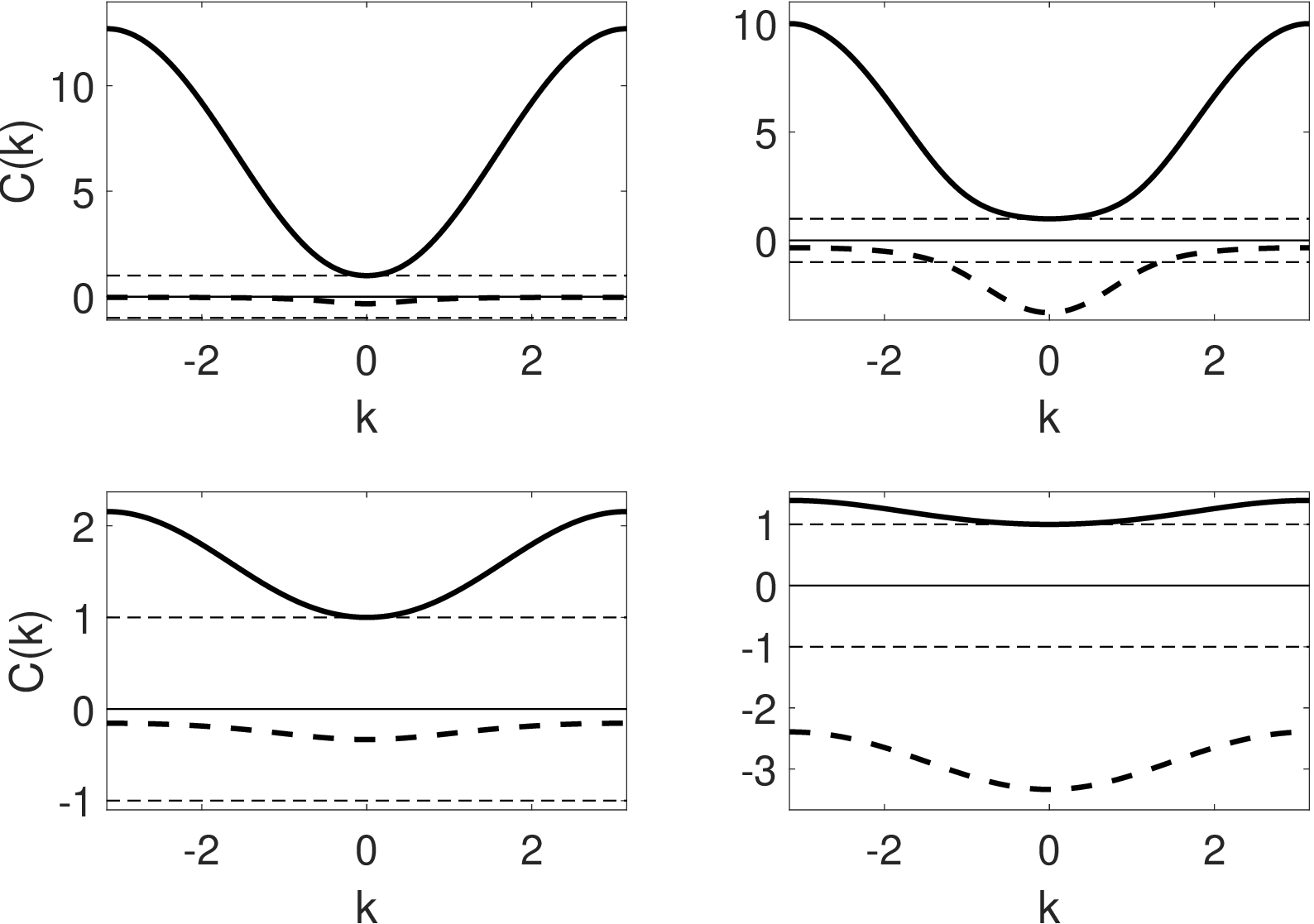} 
\caption{Illustration of the relationship between $P_1$ 
and $F_1$, namely $C(k)=P_1/F_1$ given by (\ref{CC-def}), 
for the cases  \qq{top left: $\rho=1/3$, $\mu=3$, ($m=1$, $M=3$); 
top right: $\rho=1/3$, $\mu=0.3$, ($m=10$, $M=3$); 
bottom left: $\rho=3$, $\mu=3$, ($m=1$, $M=3$); 
bottom right: $\rho=3$, $\mu=0.3$, ($m=10$, $M=3$). 
The solid lines correspond to the acoustic 
mode and the dashed lines to the optical. }
\label{CC-fig} }
\end{figure}

The solution of the matrix problem (\ref{eq11}) is given by 
\begin{align}
P_1(y,\tau,T) = & \; C(k) F_1(y,\tau,T) , \nn \\ 
C(k) = & \; \frac{\ro}{\ro-M\omega(k)^2} . 
\label{CC-def} \end{align}
The dependence of $C$ on wavenumber $k$ is
shown in Figure \ref{CC-fig}.  Note that $C$ is real 
for all wavenumbers $k$. The value of $C$ 
differs in sign between the acoustic and optical cases. 
In cases where $C<0$, the inner and outer oscillators 
are out of phase, where as $C>0$ implies the oscillators 
are in-phase.  We note that in the majority of cases 
illustrated in Figure \ref{CC-fig}, we have $|C|>1$ which 
indicates that the motion of the internal oscillators are 
larger in amplitude than the external oscillators.  The 
in-phase \qq{(acoustic)}\ modes always correspond to 
larger amplitude oscillation of the internal nodes, 
\qq{that is $C>1$};\ whereas the out-of-phase 
modes occur in both the regimes $C<-1$ and $-1<C<0$; 
the latter range corresponding to the outer oscillator 
having a larger amplitude than the inner. 
\qq{Compare the lower two panels of Figure \ref{CC-fig} 
which shows results for differing values of $\mu$; 
and note also, the top right panel, which 
shows both $C<-1$ and $-1<C<0$ depending on 
wavenumber $k$, (for the same $\mu$ and $\rho$). }\ 

The asymptotic limit cases are given by
\begin{align}
C_{\mbox{\scriptsize ac}} = &\; 
1 + 4 \mu \sin^2(\half k) ,\quad & 
C_{\mbox{\scriptsize op}} = &\; 
- \frac{1}{\mu} , 
\end{align}
for small $\mu$; and for large $\mu$ we have 
\begin{align}
C_{\mbox{\scriptsize ac}} = &\; 
1 + \frac{4}{\ro} \sin^2 (\half k) , & 
C_{\mbox{\scriptsize op}} = &\; 
- \frac{\ro}{\mu(\ro+4\sin^2(\half k))} . 
\nn \\ &&& \end{align}
Note that, whilst the relative amplitudes are $\mathcal{O}(1)$ 
in both the acoustic cases, in the optical cases, we have 
$-C_{\mbox{\scriptsize op}} \gg 1$ in the small $\mu$ limit, 
and $-C_{\mbox{\scriptsize op}}\ll1$ in the large $\mu$ limit. 

\subsection{Equations at $\mathcal{O}(\ep\ee^{0i\theta})$
\label{ep1e0-sec}}

Considering terms of $\mathcal{O}(\ep\ee^{0i\theta})$, 
we obtain the equation $0 = \rho (F_0-P_0)$ from both 
(\ref{q-eq}) and (\ref{QQ-eq}).  Thus we write $P_0=F_0$, 
where $F_0$ will be determined from higher order equations. 

\subsection{Equations at $\mathcal{O}(\ep^2\ee^{0i\theta})$
\label{ep2e0-sec}}

At this order, we again find that both equations (\ref{q-eq}) and (\ref{QQ-eq}) 
give the same relationship between $G_0,S_0,F_1,P_1$, namely 
\beq
0 =  \ro (G_0-S_0) + \ro (G_0^* - S_0^*) + 2\alpha |F_1-P_1|^2 .
\eeq
Hence, once $G_0$ is known, $S_0$ is given by 
\beq
S_0 = G_0 + \frac{\alpha}{\ro} (C-1)^2 |F_1|^2 . 
\label{S0-sol} \eeq
To determine $G_0,S_0$ independently, if these were needed, 
we would have to consider (\ref{S0-sol}) in conjunction with 
equations from the higher order, namely terms of 
$\mathcal{O}(\ep^4\ee^{0 i \theta})$ (see Sec \ref{ep4e0-sec} for details). 

\subsection{Equations at $\mathcal{O}(\ep^2\ee^{2i\theta})$
\label{ep2e2-sec}}

The second harmonic terms are governed by 
\begin{align}
\begin{pmatrix} 4 m \omega^2 \!-\! 4 \sin^2k \!-\! \rho 
& \rho \\  \rho & 4 M \omega^2 \!-\! \rho \end{pmatrix} 
\begin{pmatrix} G_2 \\ S_2 \end{pmatrix} \qquad \nn \\ 
\qquad = \begin{pmatrix} \alpha(F_1\!-\!P_1)^2 + 
16 i a F_1^2 \sin^3(\half k) \cos(\half k) \\ -\alpha(F_1\!-\!P_1)^2 
\end{pmatrix} . \nn \\ \label{e2e2eq}
\end{align}
Hence $G_2,S_2$ can be obtained from $F_1$, and $P_1=CF_1$ by  
inverting the first matrix in (\ref{e2e2eq}), which leads to 
\begin{align} 
\begin{pmatrix} G_2 \\ S_2 \end{pmatrix} = & \;  
\frac{4 F_1^2 }{ D_2 }  \begin{pmatrix} 
\alpha M \omega^2 (C\!-\!1)^2  + 4 i a \varphi (M\omega^2\!-\!\ro) \\ 
\alpha (C\!-\!1)^2 ( \sin^2 k \!-\! m \omega^2) - 4 i a \ro\varphi 
\end{pmatrix} , \nn \\ 
D_2 =&\;(4M\omega^2\!-\!\ro)(4m\omega^2\!-\!4\sin^2k\!-\!\ro)-\ro^2 ,
\nn \\ \varphi =&\;  \sin^3(\half k) \cos(\half k) . 
\label{e2e2sol} \end{align} 
Note that both $G_2$ and $S_2$ contain both real and 
imaginary components, with factors of $\alpha$ and $a$ respectively. 
The expressions 
\begin{align}
G_2 = & \; ( \hat \alpha_g + i \hat a_g ) F_1^2 , \qquad 
S_2 = ( \hat \alpha_s + i \hat a_s ) F_1^2  , \label{a-al}
\end{align}
will be used in calculations at higher order, 
to obtain a closed expression for $F_1$, see Section \ref{ep3e1-sec}, 
note that $\hat \alpha_g,\hat a_g,\hat \alpha_s,\hat a_s$ are all real.

\begin{figure}[!ht] 
\hspace*{-6mm}
\includegraphics[scale=0.31]{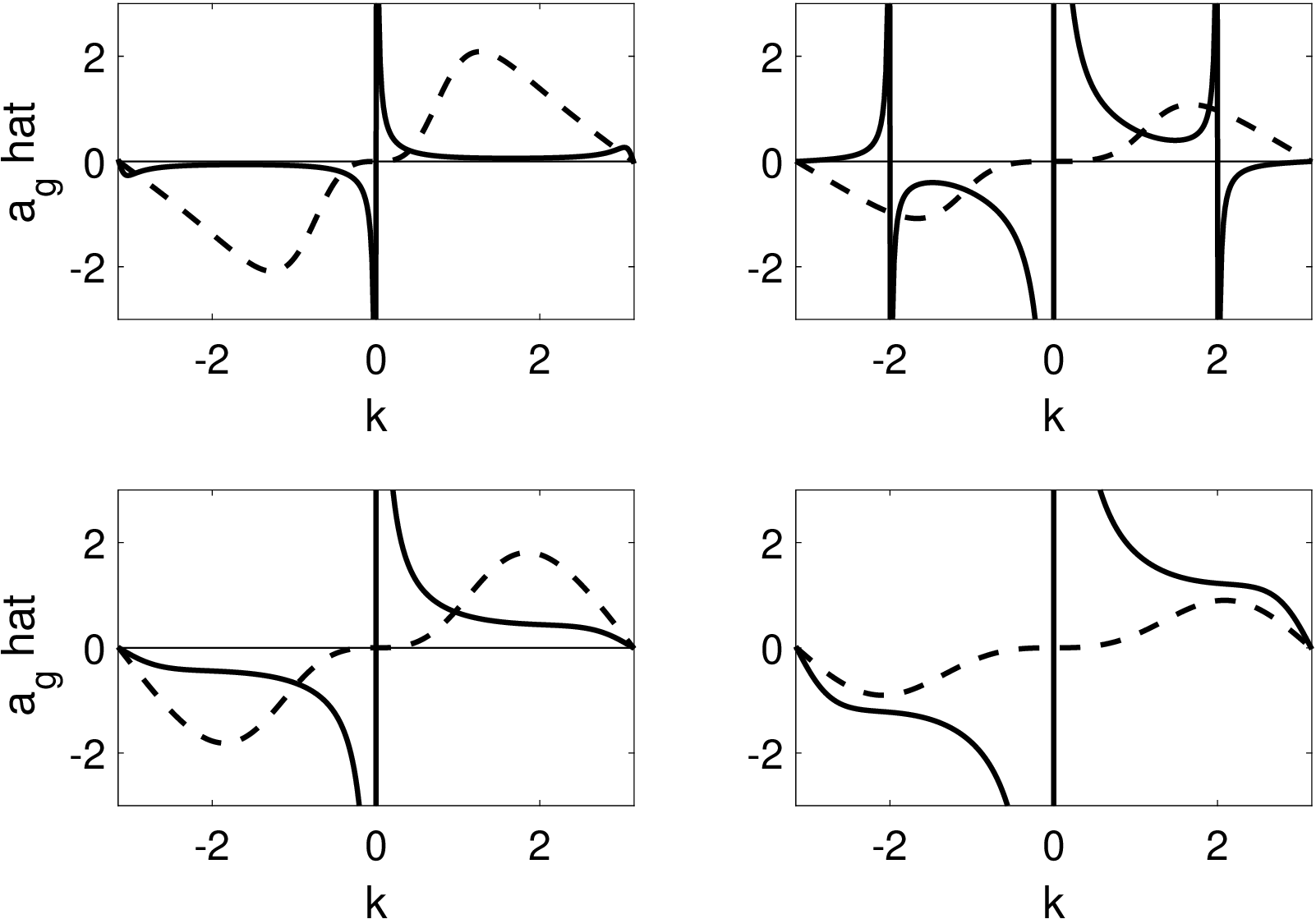} 
\caption{Illustration of the dependence of $\hat a_g$ on 
wave number $k$ given by (\ref{e2e2sol})--(\ref{a-al}), 
for the cases 
\qq{top left: $\rho=1/3$, $\mu=3$, ($m=1$, $M=3$); 
top right: $\rho=1/3$, $\mu=0.3$, ($m=10$, $M=3$); 
bottom left: $\rho=3$, $\mu=3$, ($m=1$, $M=3$); 
bottom right: $\rho=3$, $\mu=0.3$, ($m=10$, $M=3$). 
The thick solid lines correspond to the acoustic 
mode and the dashed lines to the optical}. 
In all panels, the optical cases (dashed-lines) are scaled up, 
by factors of 10,10,30,100 respectively.  
\label{ahatg-fig} }
\end{figure}

\begin{figure}[!ht] 
\hspace*{-6mm}
\includegraphics[scale=0.31]{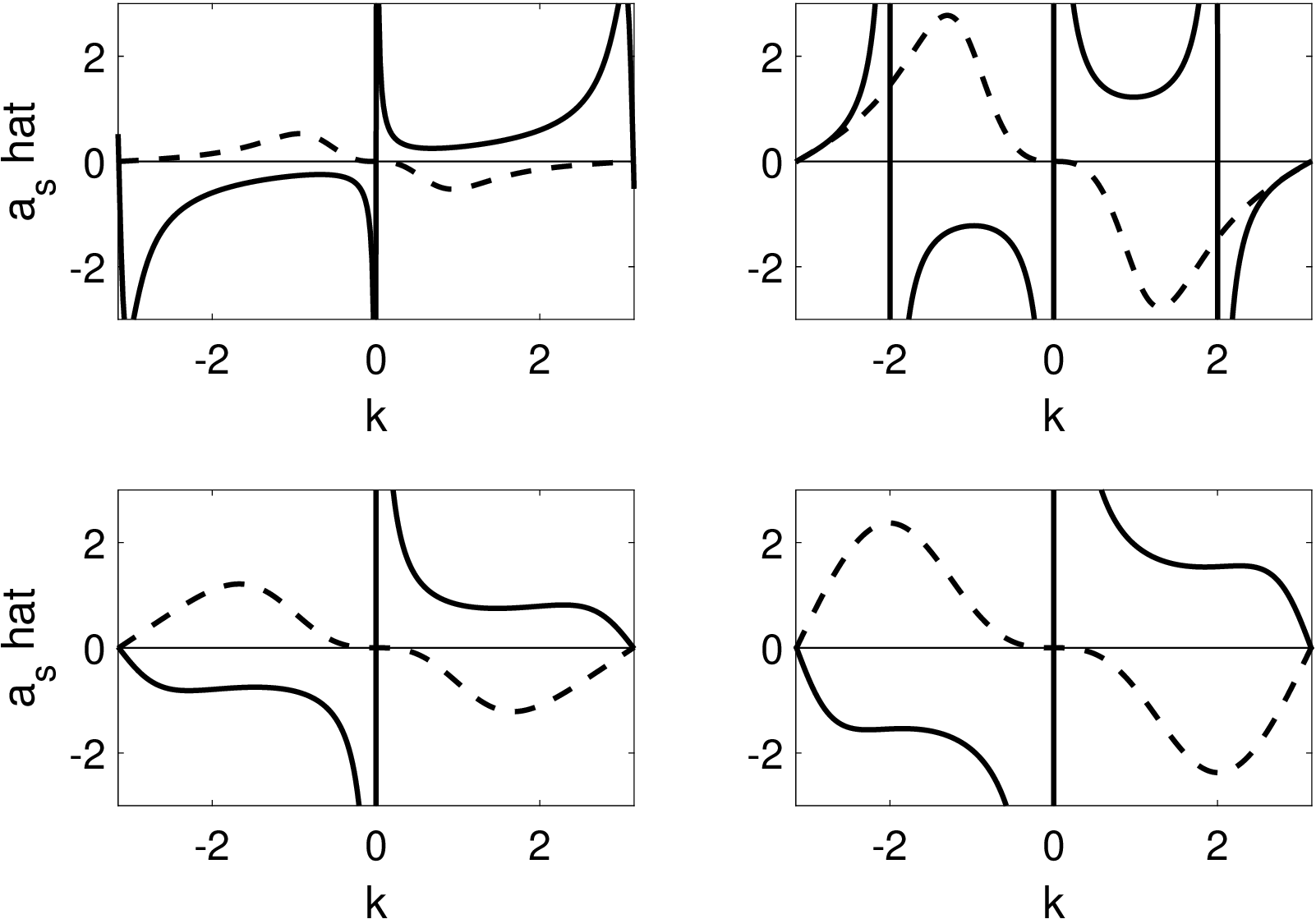} 
\caption{Illustration of the  dependence of $\hat a_s$ on 
wave number $k$ given by (\ref{e2e2sol})--(\ref{a-al}), 
for the cases  \qq{top left: $\rho=1/3$, $\mu=3$, ($m=1$, $M=3$); 
top right: $\rho=1/3$, $\mu=0.3$, ($m=10$, $M=3$); 
bottom left: $\rho=3$, $\mu=3$, ($m=1$, $M=3$); 
bottom right: $\rho=3$, $\mu=0.3$, ($m=10$, $M=3$). 
The thick solid lines correspond to the acoustic 
mode and the dashed lines to the optical}. 
In all panels, the optical cases (dashed-lines) are scaled up, 
by factors of 30,30,100,100 respectively.  
\label{ahats-fig} }
\end{figure}

\begin{figure}[!ht] 
\hspace*{-6mm}
\includegraphics[scale=0.31]{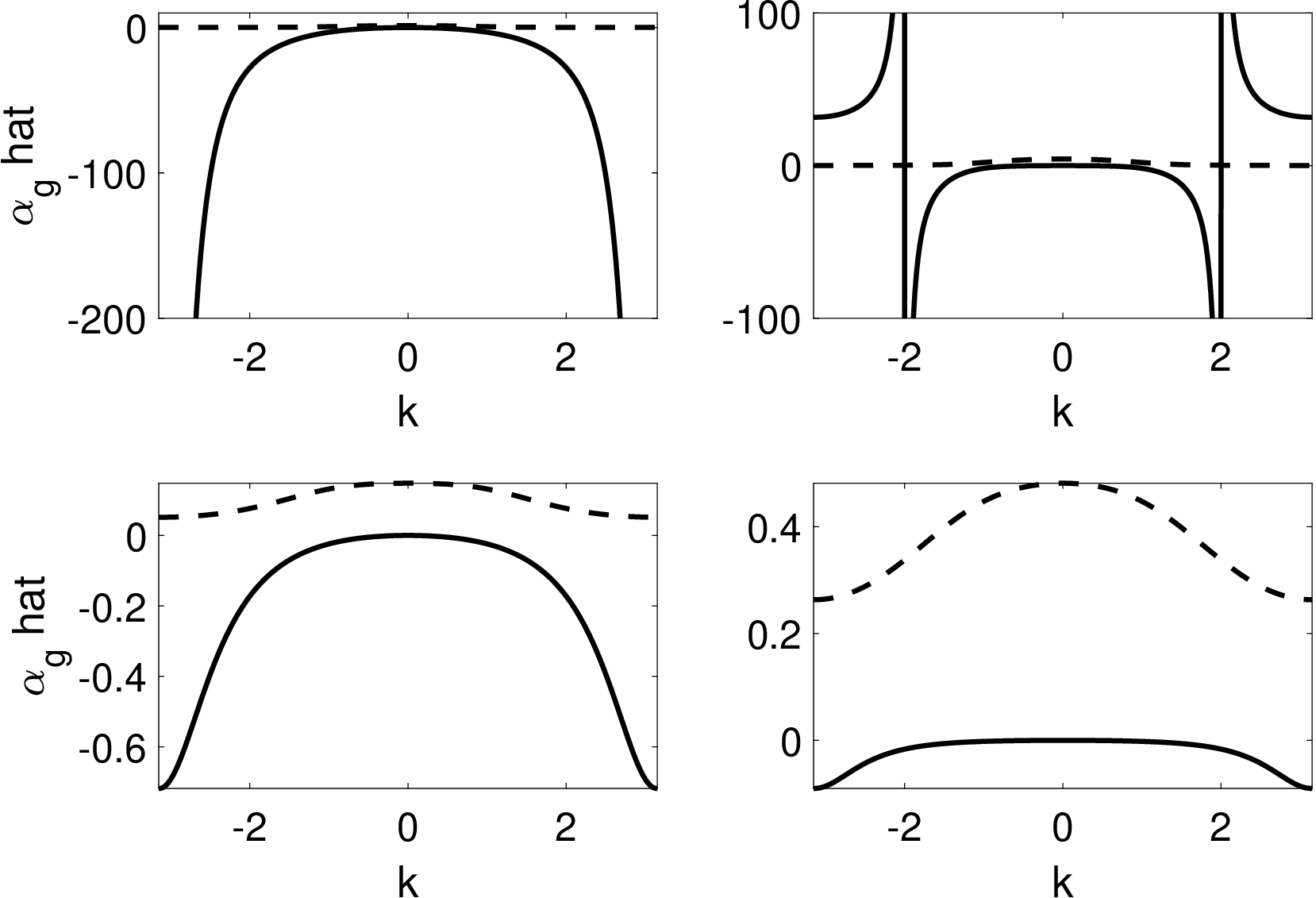} 
\caption{Illustration of the dependence of $\hat \alpha_g$ on 
wave number $k$ given by (\ref{e2e2sol})--(\ref{a-al}), 
for the cases  \qq{top left: $\rho=1/3$, $\mu=3$, ($m=1$, $M=3$); 
top right: $\rho=1/3$, $\mu=0.3$, ($m=10$, $M=3$); 
bottom left: $\rho=3$, $\mu=3$, ($m=1$, $M=3$); 
bottom right: $\rho=3$, $\mu=0.3$, ($m=10$, $M=3$). 
The thick solid lines correspond to the acoustic 
mode and the dashed lines to the optical}. 
\label{alhatg-fig} }
\end{figure}

\begin{figure}[!ht] 
\hspace*{-8mm}
\includegraphics[scale=0.32]{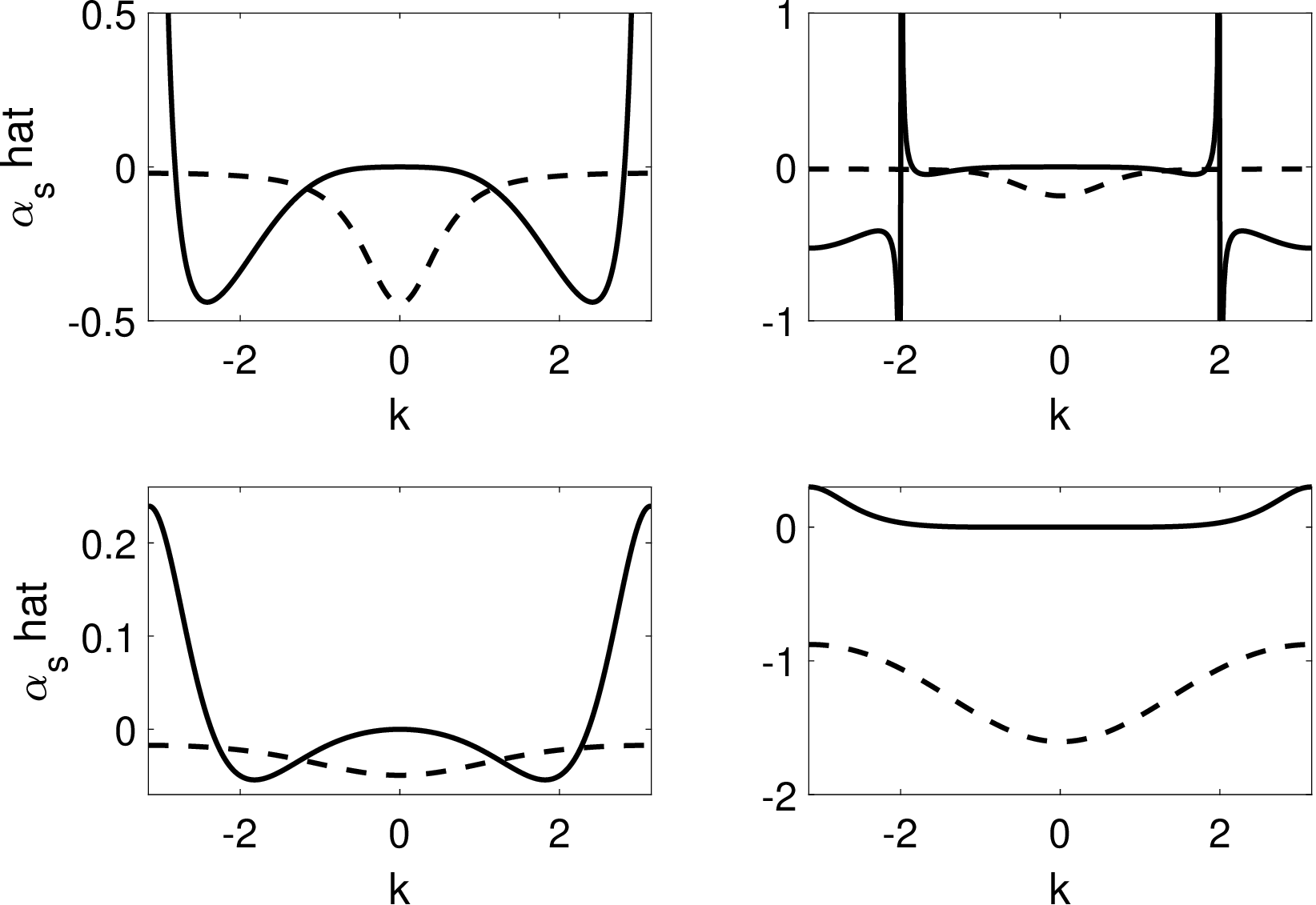} 
\caption{Illustration of the dependence of $\hat \alpha_s$ on 
wave number $k$ given by (\ref{e2e2sol})--(\ref{a-al}), 
for the cases:  \qq{top left: $\rho=1/3$, $\mu=3$, ($m=1$, $M=3$); 
top right: $\rho=1/3$, $\mu=0.3$, ($m=10$, $M=3$); 
bottom left: $\rho=3$, $\mu=3$, ($m=1$, $M=3$); 
bottom right: $\rho=3$, $\mu=0.3$, ($m=10$, $M=3$). 
The thick solid lines correspond to the acoustic 
mode and the dashed lines to the optical}. 
In the top two panels, the acoustic curves are reduced in magnitude 
by a factor of $1/200$. 
\label{alhats-fig} }
\end{figure}

Figures \ref{ahatg-fig}, \ref{ahats-fig}, \ref{alhatg-fig}, \ref{alhats-fig} 
illustrate the amplitude of the terms $\hat \alpha_g$, $\hat a_g$, 
$\hat \alpha_s$, $\hat a_s$ as functions of wavenumber $k$. 
We note that in many cases, the limit $k\rightarrow0$ in the 
acoustic case leads to a singularity.  This limit corresponds to 
the formation of a travelling wave, rather than a breather-mode, 
and different asymptotic scalings are required to consider this 
case, further details regarding travelling waves are given in 
appendix \ref{appA-sec}.  \qq{Other singularities occur 
when $D_2=0$, these happen when the frequency $\omega$ 
(\ref{disp}) satisfies 
\beq
0 = 4 \mu (m\omega^2)^2 - m\omega^2 ( \ro + \ro\mu 
+ 4 \mu \sin^2 k ) + \ro \sin^2 (k) , 
\eeq
which correspond to resonances between 
the fundamental mode and second harmonics}. 

\subsection{Equations at $\mathcal{O}(\ep^2\ee^{i\theta})$
\label{ep2e1-sec}}

The final terms at $\mathcal{O}(\ep^2)$ are those 
that have the same wavenumber and frequency as the 
leading order terms ($\ee^{i\theta}$), namely 
\beq
\begin{pmatrix} m \omega^2 \!-\! 4 \sin^2\half k \!-\! \rho & \rho \\
\rho & M \omega^2  \!-\! \rho  \end{pmatrix}
\begin{pmatrix} G_1 \\ S_1 \end{pmatrix} = {\bf b}_2 , 
\label{ep2e1eq1} \eeq
\beq
{\bf b}_2 = 
\begin{pmatrix} 
-2i \omega m F_{1,\tau} - 2 i F_{1,y} \sin k
 \\ -2 i M \omega P_{1,\tau} \end{pmatrix} . 
\label{ep2e1eq2} \eeq
This equation has the same matrix (${\bf M}$) on the {\sc lhs} 
as in (\ref{eq11}), it maps all space onto the line 
$(\ro,M\omega^2-\ro)^T$, which is the range of ${\bf M}$.  
Since ${\bf M}$ is singular, there is a Fredholm consistency 
condition on the {\sc rhs} of (\ref{ep2e1eq1}) which has to 
be satisfied in order for solutions to exist.  This condition is 
given by ${\bf n} \qq{\cdot} {\bf b}_2=0$, where ${\bf n} = 
(\ro-M\omega^2,\ro)^T$ is normal to the range of ${\bf M}$. 

We note that no nonlinear terms enter the equation 
${\bf n} \qq{\cdot} {\bf b}_2=0$ or the equation (\ref{ep2e1eq1}) 
for $(G_1,S_1)$, since the quadratic terms only generate 
second and zeroth harmonics, and no terms proportional 
to $\ee^{i\theta}$.  

\begin{figure}[!ht] 
\hspace*{-8mm}
\includegraphics[scale=0.32]{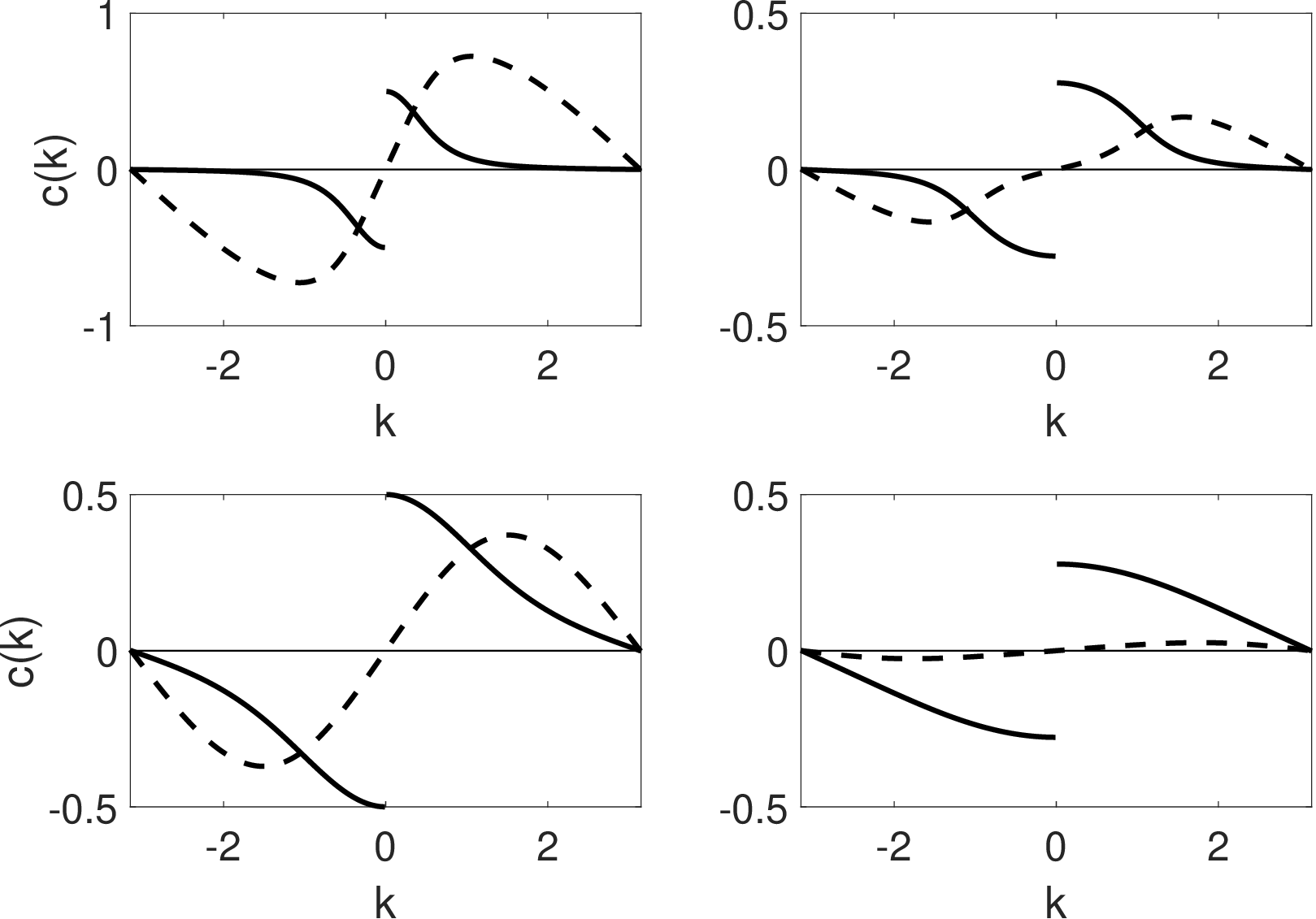} 
\caption{Illustration of the relationship between 
speed $c$ and wave number $k$  given by (\ref{c-def}), 
for the cases: 
\qq{top left: $\rho=1/3$, $\mu=3$, ($m=1$, $M=3$); 
top right: $\rho=1/3$, $\mu=0.3$, ($m=10$, $M=3$); 
bottom left: $\rho=3$, $\mu=3$, ($m=1$, $M=3$); 
bottom right: $\rho=3$, $\mu=0.3$, ($m=10$, $M=3$). 
The solid lines correspond to the acoustic 
mode and the dashed lines to the optical}. 
\label{c-fig} }
\end{figure}

Solving the \qq{consistency}\ condition ${\bf n} \qq{\cdot} {\bf b}_2=0$   
using $P_1=CF_1$, we obtain 
\beq
(\ro - M \omega^2) ( m \omega F_{1,\tau} + F_{1,y} \sin k ) 
+ \ro \omega M C F_{1,\tau} = 0 , 
\eeq
which is a first-order {\sc pde}, with a travelling wave 
solution. We write this as $F_1(y,\tau,T) = F_1(z,T)$ where 
$z=y-c\tau$ and the speed $c(k)$ is given by 
\begin{align}
c(k) = \frac{ (\rho - M \omega^2) \sin k }
{ \omega [ \ro M C \!+\! \ro m \!-\! Mm \omega^2]} = 
\frac{\sin k}{m \omega (1 \!+\! \mu C^2)} ,
\label{c-def}  \end{align}
the simplification being given by (\ref{mu-def}) and (\ref{CC-def}). 
The range of values taken by the velocity, $c$, are shown 
in Figure \ref{c-fig}.   Note that different values of the 
velocity, $c$, are obtained for the acoustic and optical cases. 
We note that the acoustic case is not well-defined 
for $k=0$, which corresponds to the case of pure travelling 
waves, as noted earlier and detailed in Appendix \ref{appA-sec}. 
Both velocities are zero when the wavenumber 
$k=\pi$, and for optical case when $k=0$. 
From hereon, we work in the moving coordinate frame, taking 
the independent variables to be $z:=y-c(k)\tau$, and $T$. 

In the limits of small $\mu$ we find the asymptotic limits  
\begin{align}
c_{\mbox{\scriptsize ac}} = &\; 
\frac{1}{\sqrt{m}} \cos (\half k) (1\!-\!\half \mu) , 
\nn \\  
c_{\mbox{\scriptsize op}} = & \; \frac{\mu^{3/2}\sin(k)}
{\sqrt{\ro m}} (1\!+\!\mfrac{3}{2}\mu) , 
\end{align}
whilst for large $\mu$, we have 
\begin{align}
c_{\mbox{\scriptsize ac}} = &\; 
\sqrt{\frac{ \mu}{m}} \cos(\half k)  \sqrt{ 1 \!+\! 
\frac{4}{\ro} \sin^2(\half k) } \, ( 1 + \mathcal{O}(\mu) ) , 
\nn \\ 
c_{\mbox{\scriptsize op}} = &\; 
\frac{ \mu \, \sin(k)}{\sqrt{m\ro}} \left(1\!+\!\frac{4}{\ro}\sin^2(\half k) 
\right)^{3/2} \, (1 + \mathcal{O}(\mu)) . 
\end{align}

\begin{figure}[!ht] 
\hspace*{-5mm}
\includegraphics[scale=0.31]{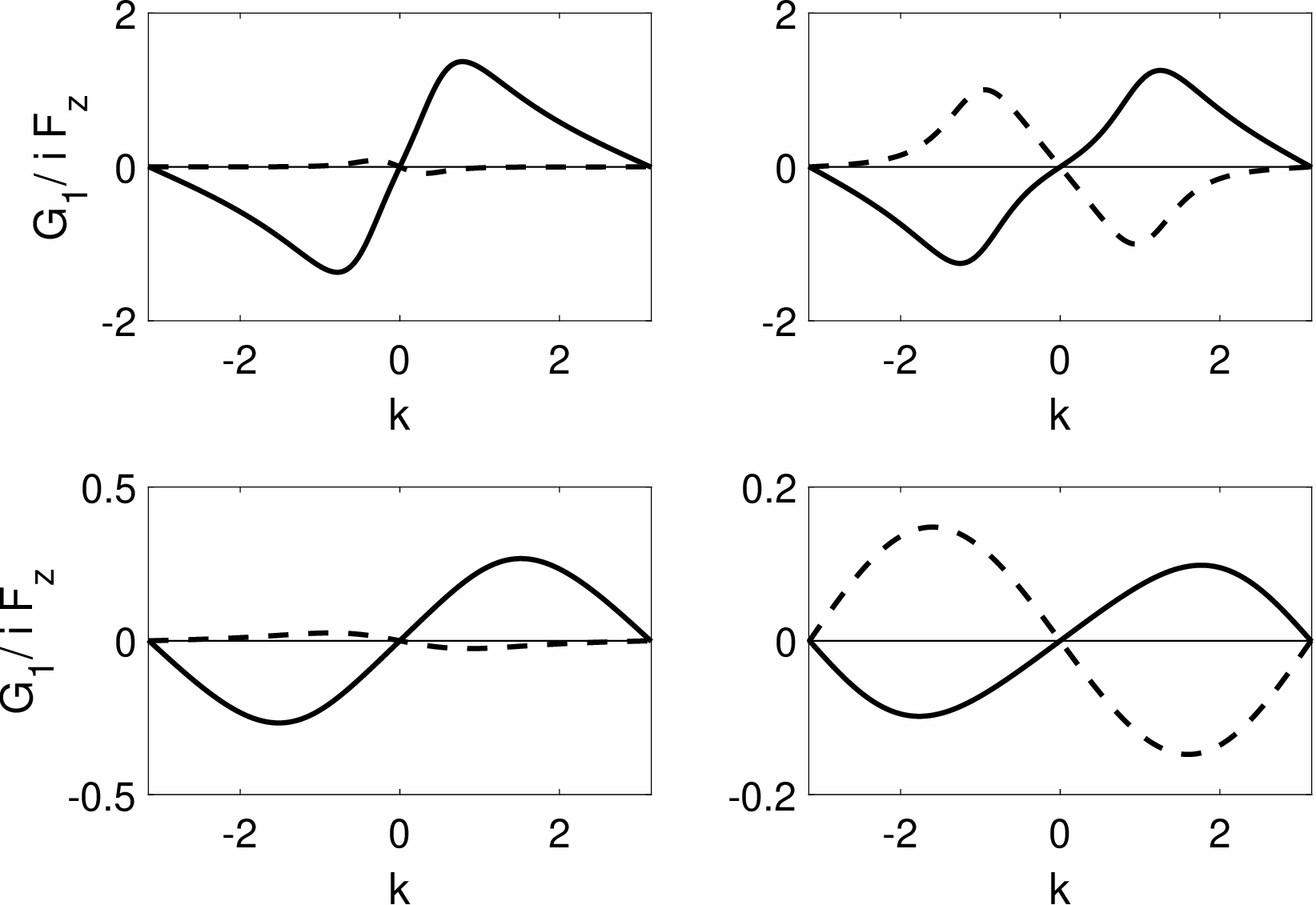} 
\caption{Illustration of the relationship between $G_1/iF_{1,z}$ 
and $k$, given by (\ref{Gtilde-sol}), for the 
cases: \qq{top left: $\rho=1/3$, $\mu=3$, ($m=1$, $M=3$); 
top right: $\rho=1/3$, $\mu=0.3$, ($m=10$, $M=3$); 
bottom left: $\rho=3$, $\mu=3$, ($m=1$, $M=3$); 
bottom right: $\rho=3$, $\mu=0.3$, ($m=10$, $M=3$). 
The solid lines correspond to the acoustic 
mode and the dashed lines to the optical}. 
\label{gam1-fig} }
\end{figure}

\begin{figure}[!ht] 
\hspace*{-5mm}
\includegraphics[scale=0.31]{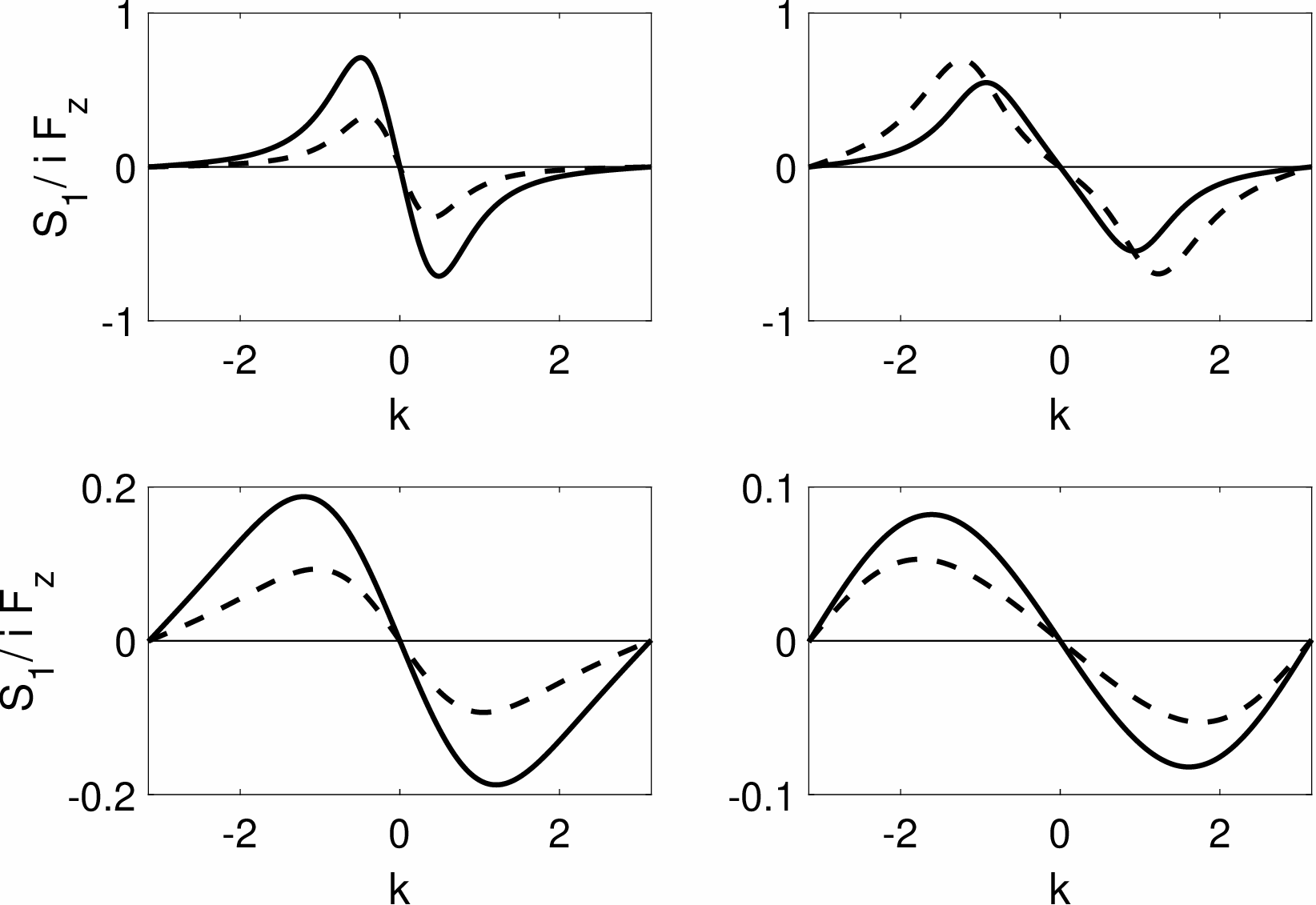} 
\caption{Illustration of the relationship between $S_1/iF_{1,z}$ 
and $k$, given by (\ref{Gtilde-sol}), for the cases: 
\qq{top left: $\rho=1/3$, $\mu=3$, ($m=1$, $M=3$); 
top right: $\rho=1/3$, $\mu=0.3$, ($m=10$, $M=3$); 
bottom left: $\rho=3$, $\mu=3$, ($m=1$, $M=3$); 
bottom right: $\rho=3$, $\mu=0.3$, ($m=10$, $M=3$). 
The solid lines correspond to the acoustic 
mode and the dashed lines to the optical.}
\label{gam1-fig2} }
\end{figure}

As well as the speed of the envelope, we need to 
determine the shape of the wave, that is, find solutions 
for $G_1,S_1$ from (\ref{ep2e1eq1})--(\ref{ep2e1eq2}). 
To solve this singular system of equations, we write 
\beq
\begin{pmatrix} G_1 \\ S_1 \end{pmatrix} = \hat G 
\begin{pmatrix} \ro \!-\! M \omega^2 \\ \ro \end{pmatrix} + \tilde G 
\begin{pmatrix} \ro \\ M \omega^2 \!-\! \rho \end{pmatrix} , 
\label{ep2e1reform} \eeq 
in this reformulation, the unknowns $G_1(z,T)$, $S_1(z,T)$ 
are replaced by $\hat G(z,T)$, $\tilde G(z,T)$. Here, $\hat G$ 
is the coefficient of the kernel of the singular matrix, so 
cannot be determined, so we assume that this is accounted 
for in the leading order terms $F_1,P_1$, and we take $\hat G=0$.  
\qq{This can be justified by considering the hypothetical case 
$\hat G \neq0$.   The $\mathcal{O}(\ee^{i\theta})$ terms in the 
asymptotic series for $(q_n,Q_n)^T$ would then start 
\begin{align}
\begin{pmatrix} q_n \\ Q_n \end{pmatrix} \sim & \; 
\ep \ee^{i\theta} F_1 \begin{pmatrix} 1 \\ C \end{pmatrix} 
+ \ep^2 \ee^{i\theta} \hat G \begin{pmatrix} 1 \\ C \end{pmatrix} 
+ \ep^2 \ee^{i\theta} \tilde G \begin{pmatrix} C \\ 1 \end{pmatrix}
\nonumber \\ & \;  + \mathcal{O}(\ep^3 \ee^{i\theta} ) , 
\end{align}
where $C$ is given by (\ref{CC-def}); note that $C\neq \pm1$ so the 
vectors $(C,1)^T$ and $(1,C)^T$ are linearly independent. 
If we define $\hat F_1 = F_1 + \ep\hat G$, then $\hat F_1$ 
satisfies the same equations as $F_1$ at leading order.  Although 
definitions of higher order terms, $H_1,H_2,R_1,R_2$ etc.~may 
be modified, our expressions for $G_0,G_2,S_0,S_2,$ remain 
unchanged.   } 

The last vector in (\ref{ep2e1reform}) is perpendicular to the kernel, 
and is not in the kernel. This enables us to find $\tilde G$. 
From the second component of (\ref{ep2e1eq1})--(\ref{ep2e1eq2}), 
we find $\tilde G = i \gamma_1 F_{1,z}$ where $\gamma_1 \in \mathbb{R}$ 
is given by 
\begin{align}
\gamma_1 = & \; \frac{2\omega cCM}{\ro^2+(M\omega^2-\ro)^2} , \nn \\ 
G_1  = &\; \rho \tilde G = i\gamma_1 \ro F_{1,z} , \nn \\ 
S_1 = & \; (M\omega^2-\ro) \tilde G = i \gamma_1 (M\omega^2-\ro) F_{1,z} .  
\label{Gtilde-sol} \end{align} 
We now have expressions for $G_1$, $G_2$, $S_0$, $S_1$, $S_2$ and 
$P_1$ in terms of $F_1$ and $G_0$.  We need to go to higher order to 
find $G_0$ in terms of $F_1$ and a closed form expression for $F_1$. 

\subsection{Equations at $\mathcal{O}(\ep^3\ee^{0i\theta})$
\label{ep3e0-sec}}

\begin{figure}[!ht] 
\hspace*{-5mm}
\includegraphics[scale=0.31]{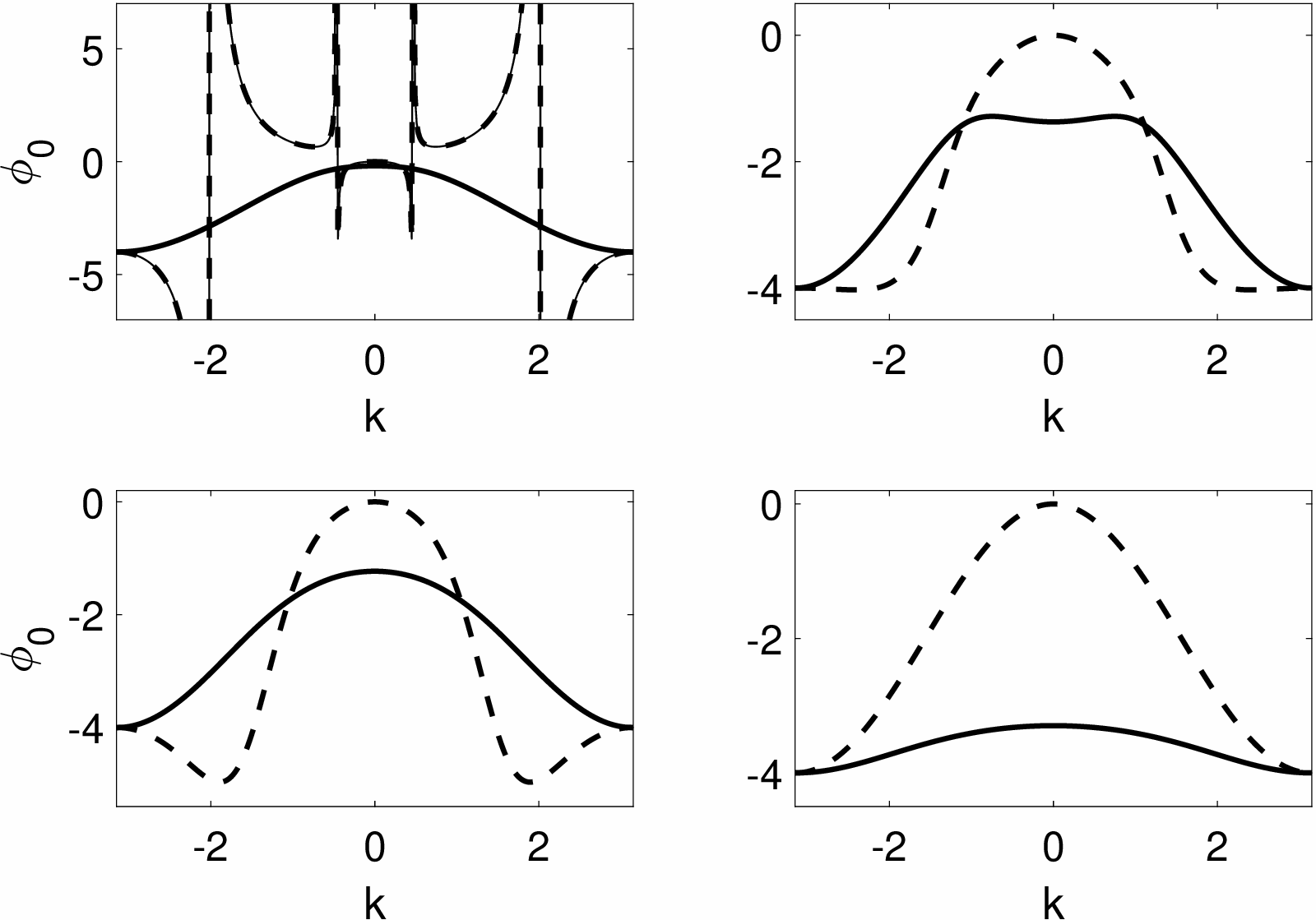} 
\caption{Illustration of $\phi_0$ against $k$, which determines 
the amplitude of the kink-shaped zero mode $F_0$ as given 
by (\ref{Fo-sol}), for the cases: 
\qq{top left: $\rho=1/3$, $\mu=3$, ($m=1$, $M=3$); 
top right: $\rho=1/3$, $\mu=0.3$, ($m=10$, $M=3$); 
bottom left: $\rho=3$, $\mu=3$, ($m=1$, $M=3$); 
bottom right: $\rho=3$, $\mu=0.3$, ($m=10$, $M=3$). 
The solid lines correspond to the acoustic 
mode and the dashed lines to the optical.}
\label{phio-fig} } 
\end{figure}

From terms of this order, we obtain the equations
\begin{align}
m F_{0,\tau\tau} \!+\! m F^*_{0,\tau\tau} =&\; F_{0,yy} + F^*_{0,yy} 
+ \ro (R_0\!-\!H_0) \nn \\ & 
+ \ro (R_0^*\!-\!H_0^*) 
 +8a \sin^2(\half k) ( |F_1|^2 )_y    \nn \\ & 
- 2 \alpha (F_1\!-\!P_1)(G_1^*\!-\!S_1^*) \nn \\ & 
- 2 \alpha (F_1^*\!-\!P_1^*)(G_1\!-\!S_1) ,  \; \label{ep3e0eq1} \\
\mu m P_{0,\tau\tau} + \mu m P^*_{0,\tau\tau} =&\; 
\ro (H_0\!-\!R_0) + \ro (H_0^*\!-\!R_0^*) \nn \\ & 
+ 2 \alpha (F_1\!-\!P_1)(G_1^*\!-\!S_1^*) \nn \\ & 
+ 2 \alpha (F_1^*\!-\!P_1^*)(G_1\!-\!S_1).
\label{ep3e0eq2}\end{align}
Noting that $P_0=F_0$, we further simplify the solution of
this system by adding the two equations together, 
transforming  to the travelling wave coordinate $z=y-c\tau$ 
with $c$ given by (\ref{c-def}). After integrating once, we find 
\beq
 F_{0,z} = a \phi_0 |F_1|^2 = 
 \frac{4 a |F_1|^2 \sin^2 (\half k)}{(1+\mu) m c^2  - 1}  . 
\label{Fo-presol}
\eeq
This represents the zero mode which gives the same 
displacements for both the inner and outer masses. 
The amplitude factor $\phi_0$ is plotted as a function of 
wavenumber $k$ in Figure \ref{phio-fig}. 

If, as frequently occurs in this type of expansion, and as will be seen 
in Section \ref{ep3e1-sec}, the equation for $F_1$ has the form of a 
nonlinear Schrodinger equation, then a typical solution has the 
form (\ref{nls-sol}), which would imply $F_0$ is given by 
\beq
F_ 0 = \frac{4 A a \sin^2(\half k) }{(1+\mu) m c^2 - 1} 
\sqrt{\frac{2D_3}{\eta}} \tanh \left( A Z \sqrt{\frac{\eta}{2D_3}} 
\right) . \label{Fo-sol}
\eeq
Thus we see the amplitude of the kink diverges near the 
speed of sound in the lattice, \qq{$c=c_0$},  (\ref{co-def}).  
\qq{From (\ref{Fo-presol}), we note that these divergences 
occur whenever the envelope wave speed $c(k)$, which is 
determined given by (\ref{c-def}) and illustrated in Figure 
\ref{c-fig} satisfies $c(k) = c_0$, where $c_0$ is the speed 
of sound of the lattice given by (\ref{co-def}). 
These divergences could be described as due to 
resonances with linear waves in the sonic limit. }

\subsection{Equations at $\mathcal{O}(\ep^4\ee^{0i\theta})$
\label{ep4e0-sec}}

Since we need to determine $G_0,S_0$ in terms of $F_1$ 
before obtaining an equation for $F_1$, we now consider 
the terms at $\mathcal{O}(\ep^4 \ee^{0i\theta}\qq{)}$ even though 
this is out of order.  We find 
\begin{align}
m (G_{0,\tau\tau} \!+\! G_{0,\tau\tau}^*) =&\; 
G_{0,yy}  + G_{0,yy}^*  
\nn \\ & 
+ 2 a i \sin(k) ( F_1 F_{1,yy}^* - F_1^* F_{1,yy}) 
\nn \\ &  + 8 a \sin^2(\half k) ( F_1^* G_{1,y} + F_1 G^*_{1,y} )
\nn \\ & 
+ \ro (U_0-I_0) + \ro (U_0^*-I_0^*)  \nn \\ & 
- 2 \alpha (F_1^*-P_1^*)(H_1-R_1)  \nn \\ & 
- 2 \alpha (F_1-P_1)(H_1^*-R_1^*) \nn \\ & 
- 2 \alpha (G_2-S_2)(G_2^*-S_2^*) \nn \\ & 
- 2 \alpha (G_1-S_1)(G_1^*-S_1^*) \nn \\ & 
- 2 \alpha (G_0-S_0)(G_0^*-S_0^*) \nn \\ & 
- \alpha (G_0-S_0)^2 
- \alpha (G_0^*-S_0^*)^2 \nn \\ & 
- 3 \beta (G_2-S_2)(F_1^*-P_1^*)^2\nn \\ & 
- 3 \beta (G_2^*-S_2^*)(F_1-P_1)^2\nn \\ & 
- 6 \beta (G_0-S_0)|F_1-P_1|^2 \nn \\ & 
- 6 \beta (G_0^*-S_0^*)|F_1-P_1|^2 , 
\label{ep4e0eq1} \\
\mu m (S_{0,\tau\tau} \!+\! S_{0,\tau\tau}^*) =&\; 
\ro (I_0-U_0) + \ro (I_0^*-U_0^*) 
\nn \\ & 
+ 2 \alpha (F_1^*-P_1^*)(H_1-R_1) \nn \\ & 
+ 2 \alpha (F_1-P_1)(H_1^*-R_1^*) \nn \\ & 
+ 2 \alpha (G_2-S_2)(G_2^*-S_2^*) \nn \\ & 
+ 2 \alpha (G_1-S_1)(G_1^*-S_1^*) \nn \\ & 
+ 2 \alpha (G_0-S_0)(G_0^*-S_0^*) \nn \\ & 
+ \alpha (G_0-S_0)^2 
+ \alpha (G_0^*-S_0^*)^2 \nn \\ & 
+ 3 \beta (G_2-S_2)(F_1^*-P_1^*)^2 \nn \\ & 
+ 3 \beta (G_2^*-S_2^*)(F_1-P_1)^2 \nn \\ & 
+ 6 \beta (G_0-S_0)|F_1-P_1|^2 \nn \\ &
+ 6 \beta (G_0^*-S_0^*)|F_1-P_1|^2 .  
\label{ep4e0eq2}  \end{align}
Adding these two equations together, and 
transforming to the travelling wave coordinate $z=y-c\tau$, 
and integrating once with respect to $z$ (and setting 
the constant of integration to zero), we find 
\begin{align}
& ( m c^2 G_{0} + \mu m c^2 S_{0} - G_{0} )_{z} = \nn \\ & \quad 
 a i \, \left[  \sin(k) - 4 \gamma_1 \ro \sin^2(\half k) \right] 
 ( F_1 F^*_{1,z} \!-\! F_1^* F_{1,z} ) . \nn \\ & \label{G0S0-eq3}
\end{align}

In the case $a=0$, this implies 
\begin{align}
m c^2 G_{0} + \mu m c^2 S_{0} - G_{0}  = &\; 0 , \label{G0S0-eq4}
\end{align}
which, when combined with (\ref{S0-sol}), gives expressions for the 
zeroth modes $G_0,S_0$ purely in terms of $F_1$ as 
\begin{align}
G_0 =&\; \gamma_0 |F_1|^2 = \frac{\alpha \mu m c^2 (C-1)^2 |F_1|^2}
{\ro( 1- m c^2 - \mu m c^2)} , \nn\\  
S_0 =&\; \sigma_0 |F_1|^2 = \frac{\alpha (1- m c^2) (C-1)^2 |F_1|^2 }
{\ro( 1 - mc^2 - \mu mc^2)} . 
\nn \\ & \label{G0S0-def} \end{align} 

\begin{figure}[!ht] 
\hspace*{-5mm}
\hspace*{-5mm}
\includegraphics[scale=0.32]{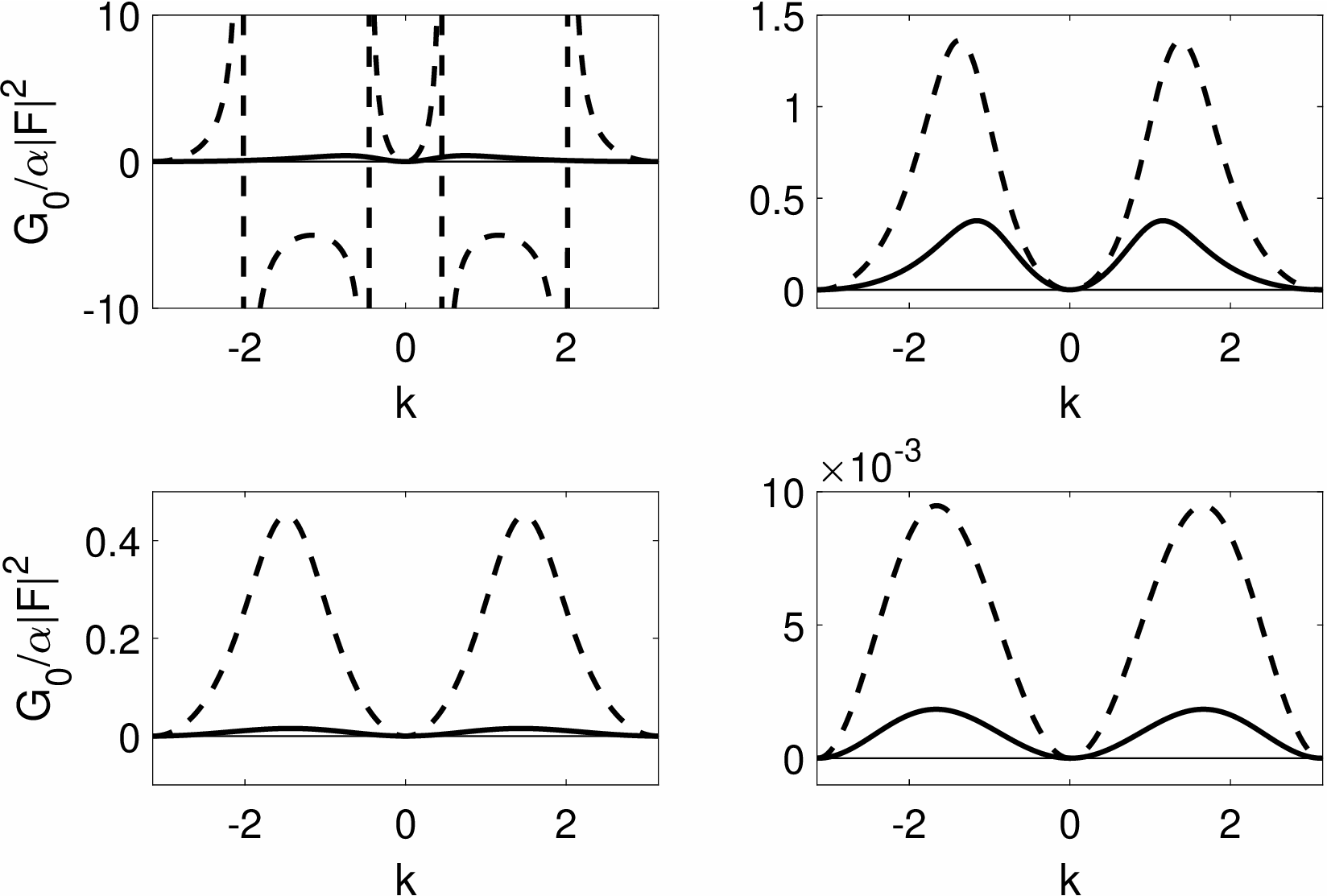} 
\caption{Illustration of the relationship between 
$\gamma_0 = G_0/\alpha|F_1|^2$ and wave number $k$  
given by (\ref{G0S0-def}), for the cases: 
\qq{top left: $\rho=1/3$, $\mu=3$, ($m=1$, $M=3$); 
top right: $\rho=1/3$, $\mu=0.3$, ($m=10$, $M=3$); 
bottom left: $\rho=3$, $\mu=3$, ($m=1$, $M=3$); 
bottom right: $\rho=3$, $\mu=0.3$, ($m=10$, $M=3$). 
The thick solid lines correspond to the acoustic 
mode and the thick dashed lines to the optical mode.}
\label{gamma-fig} }
\end{figure}

\begin{figure}[!ht] 
\hspace*{-3mm}
\includegraphics[scale=0.31]{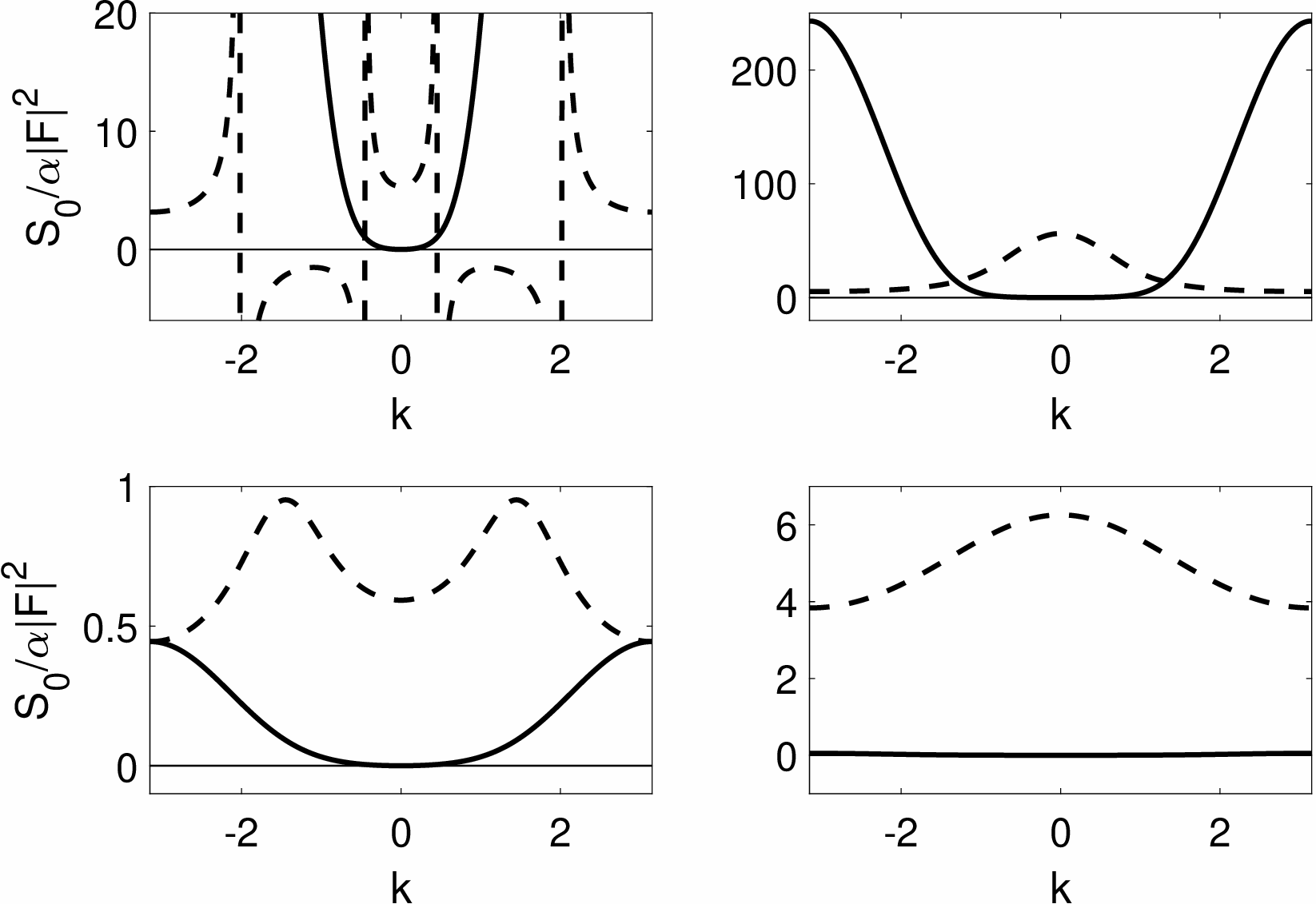} 
\caption{Illustration of the relationship between 
$\sigma_0 = S_0/\alpha|F_1|^2$ and wave number $k$  
for the cases: 
\qq{top left: $\rho=1/3$, $\mu=3$, ($m=1$, $M=3$); 
top right: $\rho=1/3$, $\mu=0.3$, ($m=10$, $M=3$); 
bottom left: $\rho=3$, $\mu=3$, ($m=1$, $M=3$); 
bottom right: $\rho=3$, $\mu=0.3$, ($m=10$, $M=3$). 
The thick solid lines correspond to the acoustic mode 
and the thick dashed lines to the 
optical mode}.  Note the wide variation 
in scales on the vertical axes.  \label{sigma-fig} }
\end{figure}

Note that this (\ref{G0S0-def}), in together with $F_0$ (\ref{Fo-sol}) 
determines the size of the zeroth harmonics.  
The $F_0$ term depends explicitly on the along-chain quadratic 
parameter $a$, and determines the leading order form of the 
zeroth harmonic,  and this component is the same for the inner 
and outer masses (since we have $P_0 = F_0$). 
The $G_0,S_0$ terms determine higher-order corrections, 
these are dependent on $\alpha$ - the coefficient of the quadratic 
nonlinearity of the potential controlling the difference in 
displacements between the inner and outer masses. 
Both these terms suffer singularities when $c^2 =1/m(1+\mu)$, 
the speed of sound in the lattice (\ref{co-def}), 
as was the case with $F_0$ (\ref{Fo-sol}).  
We plot the forms of $\gamma_0,\sigma_0$ in Figures 
\ref{gamma-fig} and \ref{sigma-fig}. 
\qq{As in Figure \ref{phio-fig}, we note that the 
graphs of $G_0,S_0$ against $k$ exhibit several 
singularities.  These occur in the same locations 
as for $\phi_0$ (\ref{Fo-presol}), and for the same reasons, 
namely that speed of the solitary wave envolope matches 
that of the speed of sound in the lattice, $c(k)=c_0$. }

In the case $a\neq0$, the solution of (\ref{G0S0-eq3}) is more 
complicated.  Writing $F_1 = J(z,T) \ee^{i \phi(z,T)}$ with $J,\phi \in 
\mathbb{R}$, we have 
\beq
i F_1 F^*_{1,z} \!-\! i F_1^* F_{1,z} = 2 J^2 \phi_z  , 
\label{Jeq1}
\eeq
so (\ref{G0S0-eq3}) expresses a relationship between real quantitites. 
If an NLS equation is obtained from the $\mathcal{O}(\ep^3 \ee^{i\theta})$ 
terms, and the solution (\ref{nls-sol}) is used for $F_1$, then $\phi_z=0$ 
and the solutions for $G_0,S_0$ given in (\ref{G0S0-def}) remain valid. 
In particular, if $F_1 = A \ee^{i \Upsilon T} J(z)$, with $A,\Upsilon,J \in 
\mathbb{R}$, then $F_1 F_{1,z}^* - F_1^* F_{1,z} = A^2 J J' - A^2 J J'=0$, 
meaning that both (\ref{Jeq1}) and the {\sc rhs} of (\ref{G0S0-eq3}) are 
zero, and so (\ref{G0S0-def}) still hold in the case $a\neq0$.

In the limit of small values of $\mu = M/m$, from 
(\ref{G0S0-def}), we find 
\begin{align}
\gamma_{0,\mbox{\scriptsize ac}}(k) \sim &\;
\frac{4 \alpha \mu^3 \sin^2(k)}{\ro} , &  
\gamma_{0,\mbox{\scriptsize op}}(k) \sim &\; 
\frac{ \alpha \mu^2 \sin^2( k) }{\ro^2} , 
\nn \\
\sigma_{0,\mbox{\scriptsize ac}}(k) \sim &\;
\frac{16 \alpha \mu^2 \sin^4(\half k)}{\ro}  , & 
\sigma_{0,\mbox{\scriptsize op}}(k) \sim & \;   
\frac{ \alpha }{\ro \mu^2} ; 
\end{align}
whilst the large $\mu$ limit gives 
\begin{align}
\gamma_{0,\mbox{\scriptsize ac}}(k) \sim &\; 
 - \frac{16\alpha \sin^4(\half k)}{\ro^3} , \nn \\ 
\gamma_{0,\mbox{\scriptsize op}}(k) \sim &\; 
 - \frac{\alpha \ro }{\mu^2 ( \ro + 4 \sin^2(\half k) )^2} , \label{goso-mubig} \\ 
\sigma_{0,\mbox{\scriptsize ac}}(k) \sim &\; 
\frac{16\alpha \sin^4(\half k) \cos^2(\half k)}{\mu \ro^3} , \quad 
\sigma_{0,\mbox{\scriptsize op}}(k) \sim \; 
\frac{ \alpha }{ \mu \ro}  . \nn 
\end{align} 
From figures \ref{gamma-fig} and \ref{sigma-fig}, we observe 
that almost all of these terms are small, the only exceptions 
being $\gamma_{0,\mbox{\scriptsize ac}} = \mathcal{O}(1)$ 
and $\sigma_{0,\mbox{\scriptsize op}} \gg1$.

\subsection{Equations at $\mathcal{O}(\ep^3\ee^{i\theta})$
\label{ep3e1-sec}}

At this final order, we obtain a system of similar form to 
Section \ref{ep2e1-sec}, but now for $H_1,R_1$, namely 
\begin{align}
\begin{pmatrix} m \omega^2 \!-\! 4 \sin^2 (\half k) \!-\! \ro & \ro \\ 
\ro & M \omega^2 \!-\! \ro \end{pmatrix}
\begin{pmatrix} H_1 \\ R_1 \end{pmatrix} = 
\begin{pmatrix} b_{31} \\ b_{32} \end{pmatrix} =: {\bf b}_3 , 
\end{align}
where 
\begin{align} 
b_{31} = &\; 
- 2 i \omega m G_{1,\tau} - 2 i \omega m F_{1,T} + m F_{1,\tau\tau} 
- 2 i  G_{1,y} \sin k \nn \\ & 
 - F_{1,yy} \cos k 
+ 2\alpha (F_1\!-\!P_1)(G_0\!-\!S_0 + G_0^* - S_0^*) 
\nn \\ & 
+ 2 \alpha ( F_1^*\!-\! P_1^*)(G_2\!-\!S_2) 
+ 3 \beta |F_1\!-\!P_1|^2(F_1\!-\!P_1) 
\nn \\ & 
- 32 a i G_2 F_1^* \sin^3(\half k) \cos(\half k) 
+ 48 b |F_1|^2 F_1 \sin^4 (\half k) \nn \\ & 
+ 8 a i F_1 (F_{0,x} + F^*_{0,x} ) \sin^2(\half k) , 
\nn \\ 
b_{32} = &\; 
-2 i \mu m \omega S_{1,\tau} - 2i \omega \mu m P_{1,T} 
+ \mu m P_{1,\tau\tau} \nn \\ & 
- 2 \alpha (F_1 \!-\! P_1)(G_0 \!-\! S_0 + G_0^*-S_0^*) 
\nn \\ & 
- 2 \alpha ( F_1^* \!-\!  P_1^*)(G_2 \!-\! S_2) 
- 3 \beta |F_1 \!-\! P_1|^2(F_1 \!-\! P_1) 
. \nn \\ & \label{b3def}
\end{align}
We do not need to solve for $H_1,R_1$, 
we only require the consistency condition on the {\sc rhs} 
for the existence of solutions, namely ${\bf b}_3 \qq{\cdot} {\bf n} = 0$, 
where ${\bf n} = (1,1+ \mu m \omega^2/(\ro-\mu m \omega^2))^T$, 
which is equivalent to the definition given after (\ref{ep2e1eq2}). 

Together with $P_1=CF_1$, and the solutions for $F_0$, 
$G_0$, $S_0$, $G_2$, $S_2$, $G_1$, $S_1$, given by 
(\ref{Fo-presol}), (\ref{CC-def}), (\ref{G0S0-def}), 
(\ref{e2e2sol}), (\ref{ep2e1reform}), (\ref{Gtilde-sol}), 
these imply 
\begin{align}
i \Omega F_{1,T} = & \; D_3 F_{1,zz} + (\eta+i \zeta) |F_1|^2 F_1 , 
\label{gnls} \end{align}
where 
\begin{align}
\Omega = & \; - \frac{2 m \omega ( \ro + \mu \ro C 
- \mu m \omega^2 ) }{(\ro - \mu m \omega^2)} , 
\nn \\ 
D_3 = & \; \frac{ \mu^2 m^2 \omega^2 c^2 C }
{(\ro - \mu m \omega^2)} + (1+\mu C)m c^2 - \cos k 
\nn \\ & 
+ 2 \ro \gamma_1 ( \sin k - (1-\mu) m \omega c ) , 
\nn \\
\eta = & \; 48 b \sin^4(\half k) + 16 a^2 \phi_0 \sin^2(\half k) 
\nn \\ & 
+ \frac{\mu m \omega^2 (C-1)}{ (\ro - \mu m \omega^2) } 
\left[ 3 \beta (C-1)^2 + 2 \alpha (\hat \alpha_g- \hat \alpha_s) 
\right. \nn \\ & \left. 
+ 4 \alpha (\gamma_0-\sigma_0) \right] 
+ 32 a \hat a_g \sin^3 (\half k) \cos (\half k) , 
\label{eta-def} \\
\zeta = & \; 32 a \hat \alpha_g \sin^3(\half k)\cos(\half k) 
- \frac{ 2 \alpha \mu m \omega^2 (1\!-\!C) 
(\hat a_g \!-\! \hat a_s)}{(\ro - \mu m \omega^2)} 
. \nn \\ & \label{zeta-def}
\end{align}

\begin{figure}[ht] 
\hspace*{-6mm}
\includegraphics[scale=0.3]{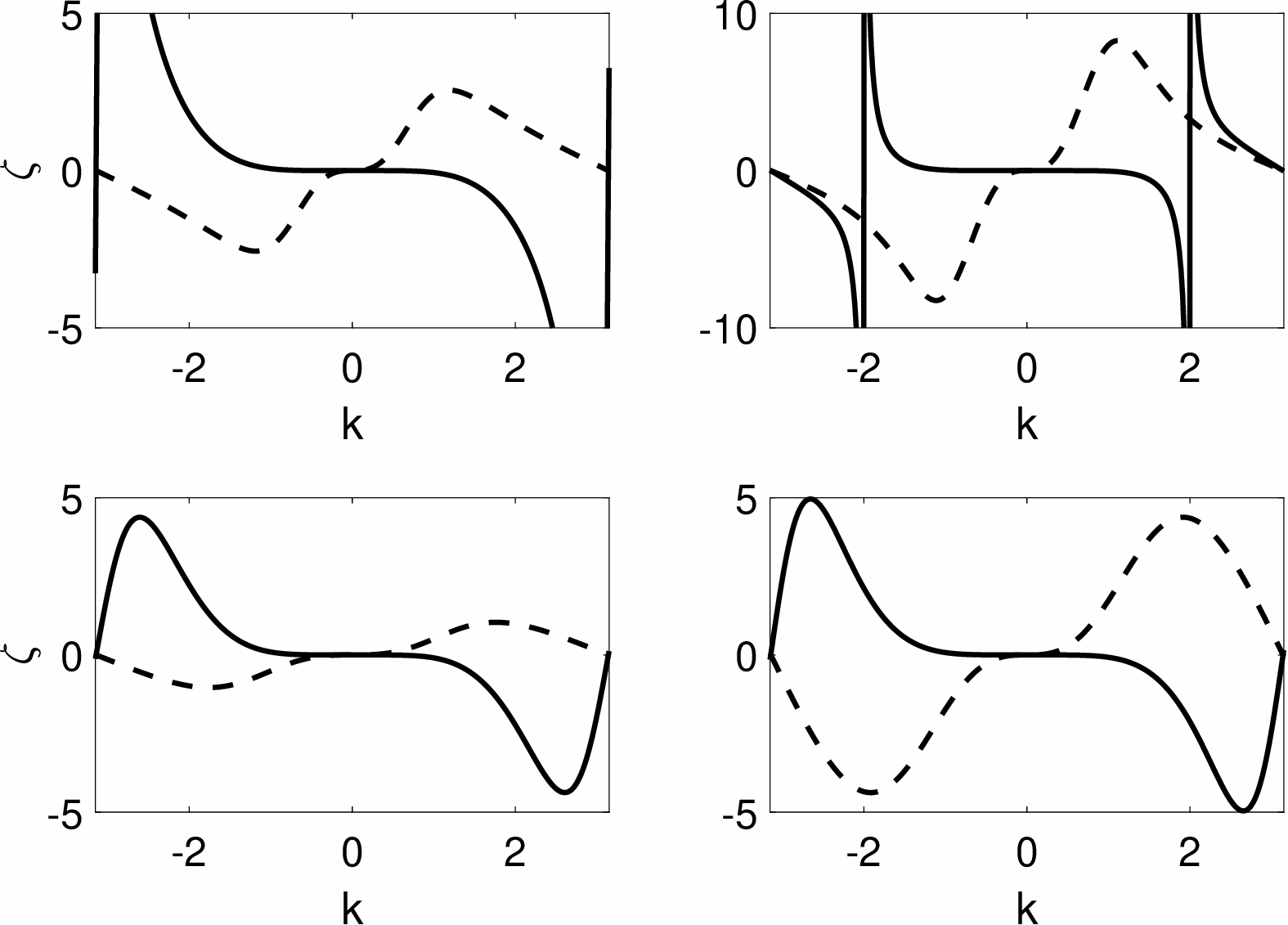} 
\caption{Illustration of $\zeta(k)$ given by (\ref{zeta-def}) plotted 
against wavenumber $k$ for the cases 
\qq{top left: $\rho=1/3$, $\mu=3$, ($m=1$, $M=3$); 
top right: $\rho=1/3$, $\mu=0.3$, ($m=10$, $M=3$); 
bottom left: $\rho=3$, $\mu=3$, ($m=1$, $M=3$); 
bottom right: $\rho=3$, $\mu=0.3$, ($m=10$, $M=3$). 
The thick solid lines correspond to the acoustic mode and 
the thick dashed lines to the optical mode}.
In all cases we take $a=\alpha=1$.   In the top two panels, 
the acoustic cases are scaled down by a factor of 200, 
in the lower right panel the acoustic case is scaled up 
by a factor of 10. 
\label{zeta-fig} } 
\end{figure}

\begin{figure}[ht] 
\hspace*{-6mm}
\includegraphics[scale=0.3]{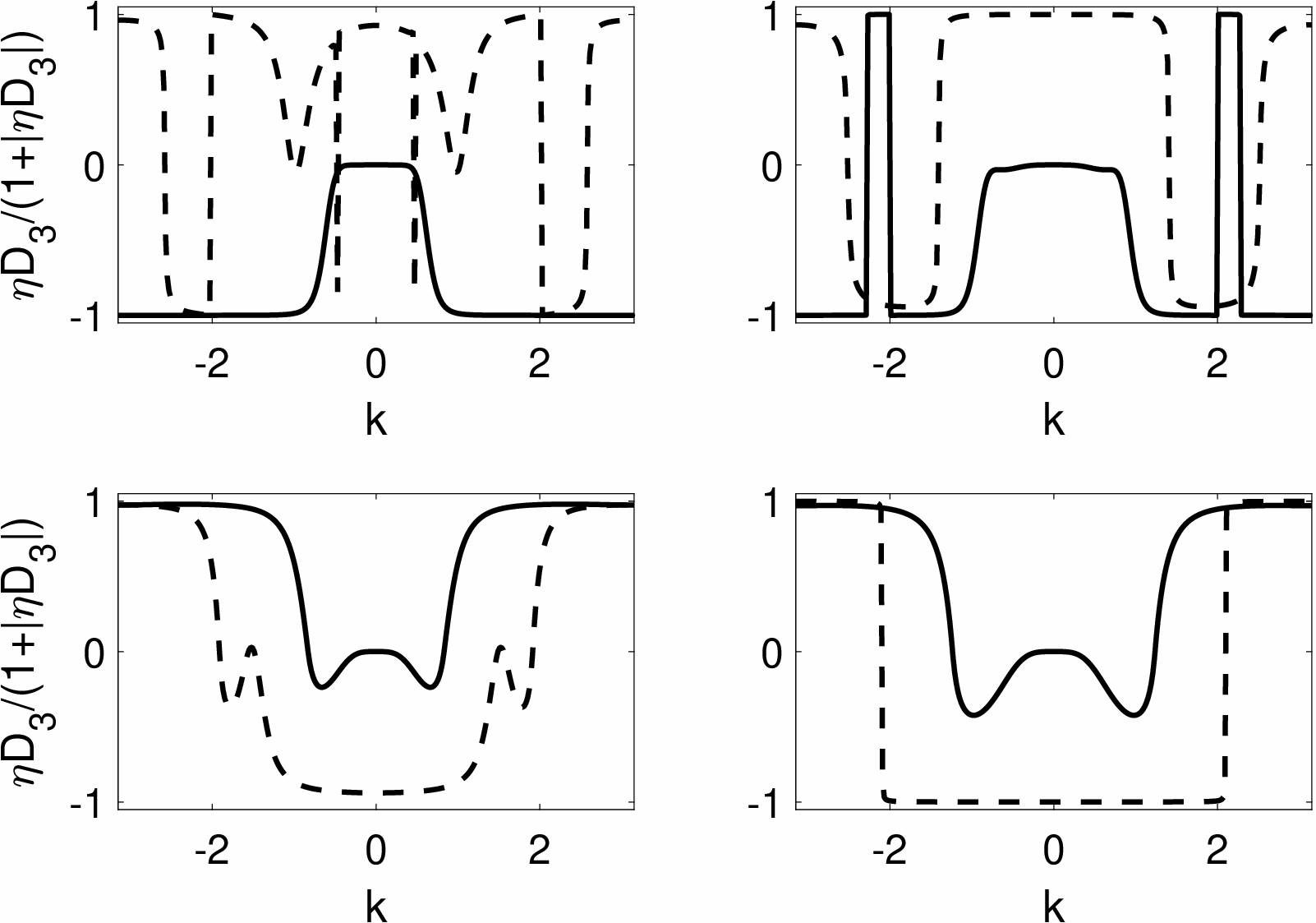} 
\caption{Illustration of wavenumbers where the NLS 
is of focusing type, that is, $\eta(k) D_3(k)>0$ given by (\ref{eta-def}).  
Due to the varying magnitude of this quantity $\eta D_3$, we plot 
$D_3(k) \eta(k) / (1 + |D_3(k)\eta(k)|)$ against wavenumber $k$ for 
the cases: 
\qq{top left: $\rho=1/3$, $\mu=3$, ($m=1$, $M=3$); 
top right: $\rho=1/3$, $\mu=0.3$, ($m=10$, $M=3$); 
bottom left: $\rho=3$, $\mu=3$, ($m=1$, $M=3$); 
bottom right: $\rho=3$, $\mu=0.3$, ($m=10$, $M=3$). 
The solid lines correspond to the acoustic 
mode and the dashed lines to the optical}.  
In all cases we take $a=\alpha=1$, $b = \beta=2$.
\label{up-fig} }
\end{figure}

In the case $\zeta=0$, the equation (\ref{gnls})  has the form of a 
nonlinear Schrodinger equation, and is of focusing form when 
$\eta D_3>0$ and defocusing form when $\eta D_3<0$. 
The range of wavenumbers where this condition is met is shown in 
Figure \ref{up-fig}.  Note that $\eta$ is dependent on 
$a,b,\alpha,\beta$, in contrast to many of the other parameters 
that have been introduced; \qq{$\eta$ also depends 
on wavenumber $k$ and linear interaction term $\ro$}. 
In particular, by increasing or decreasing 
$b,\beta$, one can change the sign of $\eta$ so that the 
condition $\eta D_3>0$ is satisfied.  
In the focusing case, the general breather solution is 
\begin{align}
F_1 = & \; A \exp\left( i K Z  +  ( 2D_3 K^2 - A^2 \eta) 
\frac{i T}{2\Omega} \right) \nn \\ &  \times  
\sech\left( \frac{A}{\Omega} \sqrt{\frac{\eta}{2D_3}} 
\left( \Omega Z +  2 K D_3 T \right) \right) .
\label{nls-gsol}
\end{align}
By absorbing the translation and spatial dependency ($K$) 
in the exponent into $c$ (\ref{c-def}), we can assume the 
simpler form (by putting $K=0$)
\beq
F_1 = A \ee^{ - i \eta A^2 T /2 \Omega} \sech \left( 
A Z \sqrt{ \frac{\eta}{2D_3} } \right) . 
\label{nls-sol}
\eeq

\qq{In cases where $\eta D_3<0$, dark breather solutions 
exist, the general form of these modes are given by 
\beq
F_1 = R(s) \ee^{iWT + i \Phi(s)} , \qquad s = z - uT , 
\label{dark-sol}
\eeq
where, following Remoissenet \cite{rembook}, $R,\Phi$ 
are determined by equating real and imaginary parts, namely 
\begin{align}
-\Omega WR+\Omega R u \Phi' =&\;\eta R^3 + D_3 R'' - D_3 R \Phi'^2 ,
\label{dark1} \\
- \Omega u R' = &\; 2 D_3 R'\Phi' + D_3 R \Phi'' . \nn
\end{align}
Integrating the latter leads to 
\begin{align}
\Phi' = & \; \frac{K}{D_3 R^2} - \frac{u\Omega}{2D_3} , 
\label{Phip-eq}
\end{align}
and substituting this into the former, (\ref{dark1}), yields 
\beq
D_3 R'' = - \Omega W R - \eta R^3 + \frac{K^2}{D_3 R^3} 
- \frac{\Omega^2 u^2 R}{4 D_3} . \label{dark2} 
\eeq
Denoting constants of integration by $K,L$, we integrate this to 
\begin{align}
(2D_3RR')^2 = &\; LR^2 - 4K^2  - (4D_3W\Omega+u^2\Omega^2) R^4
\nn \\ & - 2 D_3 \eta R^6 . 
\end{align}
The formula 
\beq
R(s) = R_0 \sqrt{1 - \nu \sech^2 \left( R_0 s 
\sqrt{\frac{-\eta \nu}{2D_3}} \right) } , 
\label{Rsol-dark}
\eeq
provides a solution under the conditions 
\begin{align}
L = & \; - 2 \eta D_3 R_0^4 (3-2\nu) , \nn \\ 
K^2 = & \; -\half \eta D_3 R_0^6 (1-\nu) , \nn \\ 
W = &\; -\displaystyle\frac{ \eta R_0^2 (3-\nu)}{2 \Omega} 
 - \displaystyle\frac{u^2 \Omega}{4 D_3}.  
\end{align}
While the first two merely assign values to the constants of 
integration, the last provides a necessary relationship for the 
wavenumber $W$ in terms of the amplitude $R_0$, speed $u$ 
and other parameters.  Integrating (\ref{Phip-eq}), we find 
\begin{align}
\Phi(s) = & \; - \frac{u \Omega s}{2 D_3} + 
R_0 s \sqrt{\frac{-\eta (1-\nu)}{2 D_3}} \nn \\ & \; + 
\tan^{\!-1} \left( \sqrt{\frac{\nu}{1\!-\!\nu}} \tanh \left( 
R_0 s \sqrt{\frac{-\nu\eta}{2 D_3}} \right) \right) , 
\end{align}
which, with (\ref{Rsol-dark}), completes the solution for 
$F_1$ (\ref{dark-sol}).   In the special case $\nu=1$ 
(where $K=0$), these equations reduce to 
$K=0$, $\Phi=-u\Omega s/(2D_3)$, 
$R = R_0 \tanh(R_0 s \sqrt{-\eta/(2D_3)})$ and hence 
\begin{align}
F_1 = \; &  R_0  \tanh \left( R_0 (z-uT) \sqrt{ \frac{- \eta}{2 D_3}} 
\right) \\ & \nn \times \exp \left( - \frac{i u\Omega}{2D_3}(z-uT) 
- i T\left( \frac{u^2 \Omega}{4 D_3} + \frac{\eta R_0^2}{\Omega} 
\right) \right) , 
\end{align}
for arbitrary $R_0,u$. This type of wave has a finite amplitude 
oscillation over all space, with a decrease in amplitude 
near $z=uT$.  }

\section{Results \label{res-sec}}
\setcounter{equation}{0} 

We consider four cases, in increasing complexity: firstly, 
Case I, where all nonlinearities are symmetric (that is, 
$a=0=\alpha$ so that $V$ and $W$, given by (\ref{V}) and 
(\ref{W}) are both even). Secondly, we consider  Case II, 
$\alpha \neq 0 = a$; thirdly, $a\neq 0=\alpha$ (Case III), and finally 
we consider the fully general Case IV where $\alpha\neq0\neq a$. 
\qq{In all cases, $\beta$ and $b$ are permitted to be nonzero, 
it is just the cases $\alpha=0$, $a=0$ which allow 
simpler results to be quoted.  The results below hold 
in the cases of $\beta=0$ and/or $b=0$, the only 
scenario which is not covered by this analysis 
is the case where $\alpha=\beta=a=b=0$. }

The case $\zeta=0$  occurs when either of the quadratic 
nonlinearities vanish, that is, $a=0$ or $\alpha=0$ or, 
\qq{when $\alpha a \neq 0$}, at  isolated values of $k$, 
such as $k=0,\pi$, as shown in Figure \qq{\ref{zeta-fig}}. 
In the case of both quadratic nonlinearities being present, 
(Case IV, $a \neq 0$ and $\alpha\neq 0$), there may be isolated 
values of $k$ where $\zeta(k)=0$; however, we might expect any 
corresponding breather solution to be unstable due to perturbations 
in the wave number causing the underlying dynamics to become 
of Ginzburg-Landau form rather than NLS. 
In the remainder of this section we consider various cases 
of $a,\alpha =0$ and $a,\alpha\neq0$  in more detail.

\begin{figure}[ht] 
\hspace*{-6mm}
\includegraphics[scale=0.42]{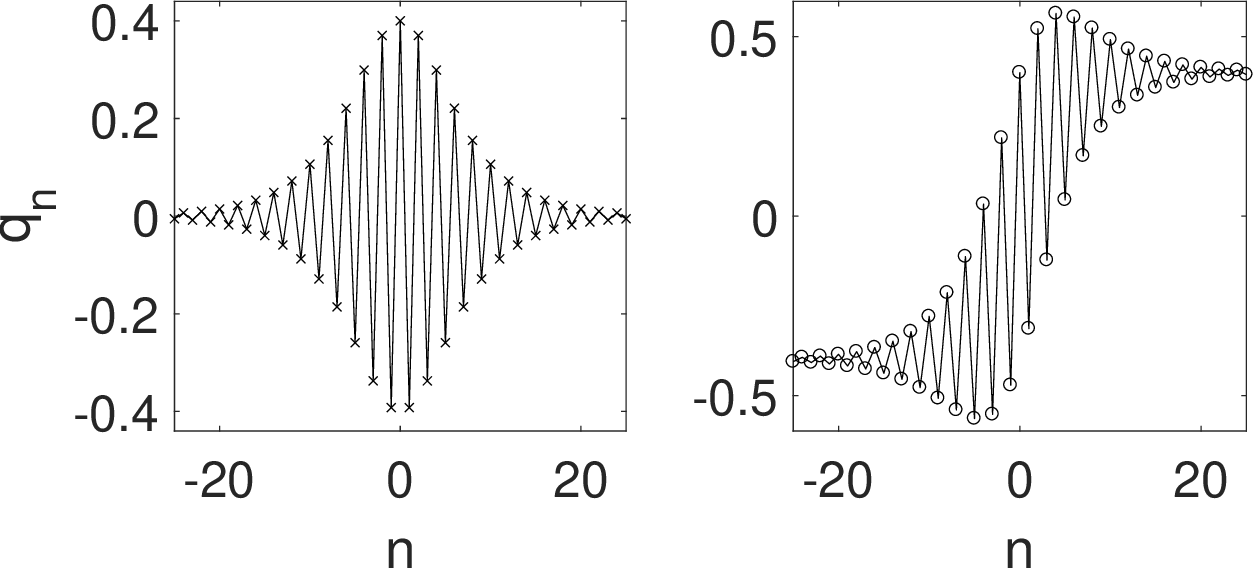} 
\caption{Illustration of the breather wave form (\ref{br-sol}) 
and breather-kink waveform (\ref{brk-sol}). 
\qq{Left: $q_n = 2 h \sech(hn) \cos( \pi n)$; 
right: $q_n = 2h \sech(hn) \cos(\pi n) + 2h \tanh(h n)$  
with $h=0.2$ in both cases.} 
\label{br-brk-fig} }
\end{figure}

\subsection{Case I: even potentials ($a=0=\alpha$) 
\label{a0a0-subsec}}

Putting $\alpha=0=a$ simplifies the problem considerably. 
The dispersion relation (\ref{disp}) remains, 
together with $P_1=CF_1$ with $C=C(k)$ given by (\ref{CC-def}). 

In this case we have $S_0=G_0=0$ from (\ref{S0-sol}) and 
(\ref{G0S0-def}), $G_2=S_2=0$ from (\ref{e2e2eq}), but we 
still have $G_1,S_1$ not necessarily zero (\ref{Gtilde-sol}). 
The envelope speed remains (\ref{c-def}), 
and first-correction terms are given by (\ref{Gtilde-sol}). 
The main simplification is that we have $\zeta(k)=0$ in 
(\ref{gnls})--(\ref{zeta-def}). 

In this case, the breather mode is simple, having no zero-mode 
contributions from $F_0$ or $G_0,S_0$. 
The leading order solution for the breather is 
\begin{align}
q_n(t) \sim & 2 \ep A \cos \left( k n - \omega(k) t 
- \frac{\ep^2 A^2 \eta(k) t}{2\Omega(k)} \right) \nn \\ & 
\times \sech \left( \ep A  (n-ct) \sqrt{\frac{\eta(k)}{2D_3(k)}}  
\right)  . 
\label{br-sol}
\end{align}
This form is illustrated in Figure \ref{br-brk-fig}.  
At leading order, we have $Q_n(t) = C(k) q_n(t)$ 
with $C(k)$ given by (\ref{CC-def}). 
Note that this solution is depends on the two parameters, 
$k$ which governs the wavenumber of the linear carrier wave, 
and the amplitude $\ep A$.  

\subsection{Case II:  $a=0$, $\alpha\neq0$ \label{ao-sec}}

Allowing the force between the inner and outer particles 
to have a quadratic component ($\alpha\neq0$, but with $\beta=0$) 
whilst that of along chain has no quadratic component ($a=0$ 
with $b \neq 0$) still results in (\ref{gnls}) being a NLS. 
The leading order breather solution is again given by (\ref{br-sol}).
In this case a zero-mode is produced, that is, $G_0,S_0\neq0$ 
by (\ref{G0S0-def}), however, this is small correction term; 
and there is no leading order zero mode, since we have $F_0=0$ 
from (\ref{Fo-presol}).  The zero mode (\ref{G0S0-def}) is 
localised to the site of the breather. 

\subsection{Case III:  $a\neq0$, $\alpha=0$}   \label{III-sec}

We now reverse the situation from Sec \ref{ao-sec}. 
We allow the along chain interactions to have a quadratic 
component to the force ($a\neq0$), whilst 
requiring the force between the inner and outer particles 
to have only a cubic nonlinearity ($\alpha=0$, $\beta\neq0$).  
This again results in (\ref{gnls}) being a NLS, 
since $\zeta=0$ for all wavenumbers $k$. 

The dispersion relation is given by (\ref{disp}), $P_1=CF_1$ 
with (\ref{CC-def}). 
Since $\alpha=0$, from (\ref{S0-sol}), we have $G_0=S_0 = 0$, 
however, a zero mode (\ref{Fo-presol}) is produced due to 
$F_0\neq0$, this mode is the same for both inner and outer 
masses.  The mode is not localised: given that $F_1$ has a 
sech-shape, $F_0$ has a tanh form, so this corresponds 
to a localised pre-compression of the lattice.   
In (\ref{e2e2sol}), there is some simplification, although 
second harmonic terms are still generated. 
From (\ref{G0S0-def}), we find $G_0=S_0=0$, 
so the only zero-mode we are concerns is due to $F_0$. 

The leading order solution (\ref{br-sol}) is replaced by the 
more general kink-breather combination 
\begin{align}
q_n(t) \sim & 2 \ep A \cos \left( k n - \omega(k) t 
- \frac{\ep^2 A^2 \eta t}{2\Omega} \right) \nn \\ & \!
\times \sech \left( \ep A  (n-ct) \sqrt{\frac{\eta}{2D_3}}  
\right) \label{brk-sol} \\ & 
\!\!+\!  \frac{8 \ep a A \sin^2(\half k)\sqrt{2D_3} }
{\sqrt{\eta} [ (M\!+\!m) c^2 -1 ]} \! \tanh \!\left( 
\frac{\ep A (n-ct) \sqrt{\eta} }{ \sqrt{ 2D_3}} \right) .  
\nn \end{align}
This form is illustrated in Figure \ref{br-brk-fig}. 

\subsection{Case IV:  the general case $a \neq 0 \neq \alpha$}
\label{fully-quad-sec} 

For $a\neq0 \neq \alpha$, equation (\ref{gnls})  is of complex 
Ginzburg-Landau form (CGL), rather than an NLS equation. 
For some values of the wavenumber $k$, 
we may have $\zeta(k)=0$ and so the NLS derivation is valid; 
however, for general values of $k$, we expect $\zeta(k) \neq 0$, 
and so different dynamics may be observed.  
When $\zeta=0$, it is natural to consider this as a combination 
of Cases II and III, (Secs \ref{ao-sec} \& \ref{III-sec}), so that 
both the leading order nonlocal zero mode, $F_0$ (which 
is the same for both inner and outer masses), and  
the smaller, localised zero modes $G_0\neq S_0$ 
are present, the latter giving different amplitudes 
for the inner and outer masses. 
For example, if we consider the special case $k=\pi$, then we find 
$c=0 = \gamma_1 = G_1 = S_1 = G_0 = \hat a_g = \hat a_s$, 
and $\zeta=0$, so the NLS reduction remains valid, and 
long-lived stationary breather-modes may be expected to exist. 

\begin{figure}
\includegraphics[scale=0.5]{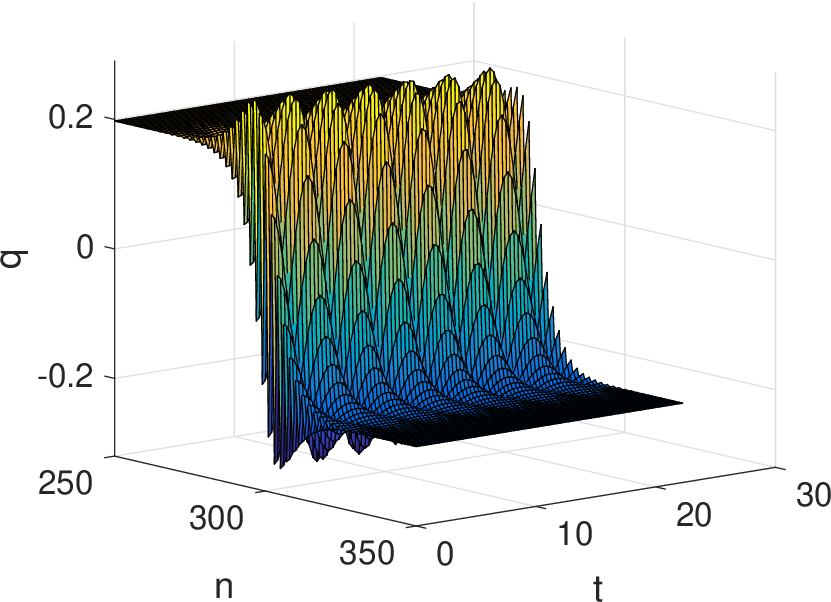} \\ 
\includegraphics[scale=0.5]{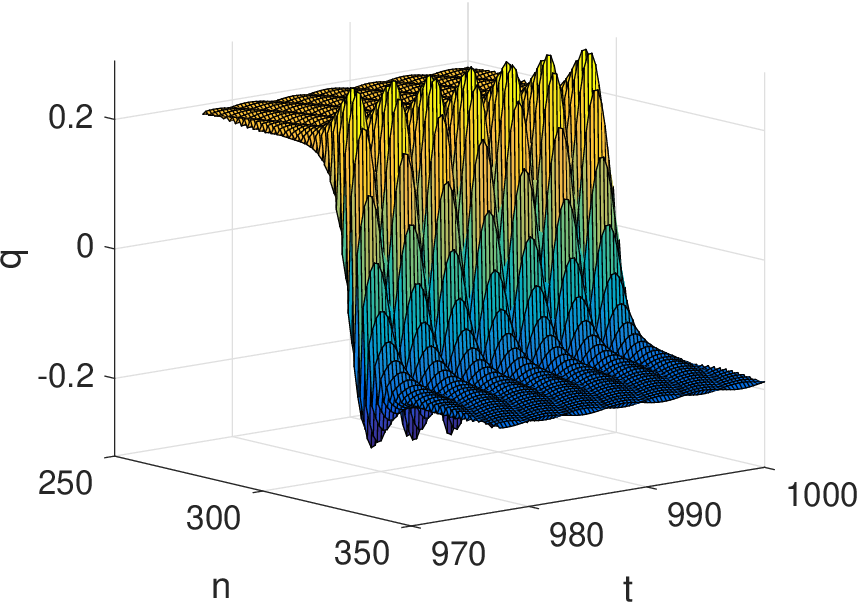}
\caption{\qq{Results of numerical simulations of the full 
ODE system. The system is simulated using Matlab ode45 
\cite{matlab} with $N=400$ lattice sites, and parameter values 
given by $\rho=3$, $m=3$, $M=1$, $\ep=0.1$, $\mu = 1/3$, 
$\ro =3$, $a=0.5$, $b=0.5$, $\beta=2$, $\alpha=0.5$, $k=\pi$. 
The plots show the resulting acoustic mode at the start and 
end of a simulation of length $t_{max}=1000$.  
In colour in online version.  } \label{sim-fig}}
\end{figure}

\qq{ In Figure \ref{sim-fig} we illustrate the results of 
a numerical simulation of the system (\ref{q-eq})--(\ref{QQ-eq}), 
started with initial conditions given by $k=\pi$ in the leading order 
terms ($F_0,F_1,P_0,P_1$) from (\ref{q-ans})--(\ref{QQ-ans}), 
namely (\ref{nls-sol}) and (\ref{Fo-presol}).    
This corresponds to Case IV, since  $\alpha\neq0\neq \beta$ 
and  $a\neq0\neq b$. 
We have neglected the second order and all higher terms 
($G_j,H_j,S_j,R_j,\ldots$); the initial conditions are an 
approximation to the mode. 
Over early times ($0<t<\sim10^2$), there is a very small 
adjustment in the shape of the mode over longer times, the 
mode appears stable.  The system has a small amount of damping 
built into lattice sites $1 \neq n \leq 40$ and $360 \leq n \leq 400$ 
in order to dampen the radiation which is shed from the mode 
at early times. This takes the form of additional terms 
$-\lambda \dd q_n/\dd t$ and $-\lambda \dd Q_n/\dd t$ added to 
equations (\ref{q-eq})--(\ref{QQ-eq}) with $\lambda=0.03$. }

When $\zeta\neq0$, (\ref{gnls}) has the form of a complex 
Ginzburg Landau equation (CGL), which exhibits  
markedly different behaviour from NLS.  
The CGL equation is typically written as 
\beq
F_{1,T} = \delta F_1 + (1 +i \chi) F_{1,XX} + 
(1+ i \zeta) |F_1|^2 F_1 . 
\eeq
Our case (\ref{gnls}) corresponds to the limit $\chi\gg 1$, 
with $X\gg1$,  $\zeta=\mathcal{O}(1)$, and $\delta=0$. 
In these cases the equation (\ref{gnls}) does not have stable 
pulse-type solutions. Instead,  solutions of (\ref{gnls}) either 
decay to zero or blow up according to the sign of $\xi$. We 
find that norm of the distribution $F_1$ evolves according to 
\begin{align}
\Omega \frac{\dd  }{\dd t} \int |F_1|^2 \, \dd z = &\;  
2 \zeta \int |F_1|^4 \, \dd z , 
\end{align}
thus if $\zeta>0$, then $\int |F_1|^2 \dd z$ monotonically 
increases and if $\zeta<0$ then $\int |F_1|^2 \dd z$ 
monotonically decreases. 
Similarly, the NLS Hamiltonian is not conserved 
when $\zeta\neq0$.  If we define 
\beq
H = \int D_3 | F_{1,z} |^2 - \half \eta |F_1|^4 \, \dd z , 
\eeq
then we find 
\begin{align}
\!\frac{\dd H}{\dd t} =&\; \frac{\zeta D_3}{\Omega} \!\int 4 |F_1 F_{1,z} |^2 
+ F_1^2 F_{1,z}^{*2} + F_1^{*2} F_{1,z}^2 - \frac{2 \eta}{D_3} |F_1|^6 \, \dd z 
. \nn \\ & 
\end{align}
A full discussion of the dynamics exhibted by the Ginzburg-Landau 
equation is beyond the scope of this paper, we refer the 
interested reader to the wider literature, for example, the 
introductions provided by Garcia-Morales \& Krischer 
\cite{GL-intro2} and Hohenberg \& Krekhov \cite{GL-intro1}.

\section{Conclusions}
\setcounter{equation}{0}

We have considered the fully nonlinear problem of a mass-in-mass FPUT 
chain in which {\em both} the along-chain interactions and the interaction 
between the inner and outer masses are nonlinear.  We have used 
multiple scales asymptotics to construct an explicit form for the breather 
in the small amplitude limit.  This involves solving systems of equations at 
each order of magnitude and for each harmonic frequency, using a 
Fredholm consistency condition to generate additional solution criteria. 
In many cases, this ultimately yields a nonlinear Schrodinger equation. 

Many asymptotic approximations of breathers require the calculation 
of the ``zero''-mode corrections at $\mathcal{O}(\ep^2 \ee^{0i\theta})$,
from equations at $\mathcal{O}(\ep^4 \ee^{0i\theta})$.   Often this 
requires knowledge of the leading order term, so giving a coupled problem.  
However, in the case analysed here, the equations at 
$\mathcal{O}(\ep^2 \ee^{0i\theta})$ and 
$\mathcal{O}(\ep^4 \ee^{0i\theta})$ give explicit formulae for $G_0,S_0$. 
This enables us to calculate a single NLS equation for the leading order 
shape, $F_1$. 

In addition to illustrating properties of the solution for various choices of 
the masses, we have given simplified asymptotic forms for the solution 
in the cases where the ratio of the inner to outer masses is extremely 
small or extremely large.   In future work \cite{reem}, we propose to use 
numerical techniques to investigate the stability, robustness, and other 
properties of breather solutions in this system, in both the cases 
$a=0$ and $a\neq0$. 

\subsection*{Acknowledgements}

I am grateful to Paul Matthews for helpful conversations regarding the 
complex Ginzburg-Landau system. 
\qq{I am grateful to Reem Almarashi for writing some of the code 
which produced Figure \ref{sim-fig}. 
My thanks also go to the referees for making helpful comments 
to improve the manuscript.} 

\appendix

\renewcommand{\theequation}{\Alph{section}\arabic{equation}}
\renewcommand{\thesubsection}{\thesection\arabic{subsection}}

\section{Asymptotics of travelling wave solutions} 
\label{appA-sec}
\setcounter{equation}{0}

Below, we derive the small amplitude travelling wave solutions 
for the fully nonlinear lattice.  The results show that the nonlinear 
nearest-neighbour interactions control the shape of the waves, 
and that the nonlinear interaction between inner and outer masses 
is only relevant in higher order terms. 
There are two separate cases to consider as 
different asymptotic scalings are required for the cubic ($a\neq 0$)
and quartic ($a=0$) nearest neighbour potential energy functions.

\subsection{Quartic potential, $a=0$, $b\neq0$}
\label{tw-pure-quart-sec}

Here we leave $\alpha,\beta$ arbitrary. The governing ODEs are 
\begin{align}
\!\!m \frac{\dd^2 q_n}{\dd t^2} \!=& q_{n+1} \!-\! 2 q_n \!+\! q_{n-1} 
 \!+\! b (q_{n+1}\!-\!q_n)^3 \!-\! b (q_n\!-\!q_{n-1})^3 
\nn \\ & - \rho (q_n-Q_n) - \alpha (q_n-Q_n)^2 - \beta (q_n-Q_n)^3 , 
\nn \\ & \\ 
\!\!\!\!M \frac{\dd^2 Q_n}{\dd t^2} =&\; \rho(q_n-Q_n) 
+ \alpha (q_n-Q_n)^2 + \beta (q_n-Q_n)^3 .  \nn \\ & 
\end{align}
We replace $(q_n(t),Q_n(t))$ by $(q(y,\tau),Q(y,\tau))$ using 
the scaled variables  
\begin{align}
& y= h  n , \qquad \tau = h t ,  \qquad h \ll 1 ,
\label{app-tw-a-s1} \\ 
& Q = q - h^{2} W , \qquad M = \mu m ,  
\label{app-tw-a-s2}
\end{align}
to obtain the approximating PDEs 
\begin{align}
m q_{\tau\tau} = & \; q_{yy} + \rec{12} h^2 q_{yyyy} 
+ b h^2 (q_y^3)_y - \rho W -\alpha h^2 W^2  , \\ 
\mu m q_{\tau\tau} = & \; \rho W + \alpha h^2 W^2 + h^2 \mu m W_{\tau\tau} , 
\end{align}
where we have neglected terms of $\mathcal{O}(h^4)$ and higher. 

If we assume a TW of the form $q(y,\tau) = q(z)$, $Q(y,\tau)=Q(z)$, 
$W(y,\tau)=W(z)$, where $z = y - c \tau$, then we obtain the system of ODEs 
\begin{align}
(m c^2 -1)q'' + \rho W = & \;  h^2 \left[ 
\rec{12} h^2 q'''' + 3b (q'^2) q'' - \alpha W^2\right] , 
\label{quart-upp} \\ 
\mu m c^2 q'' - \rho W = & \; h^2 \left[ \alpha W^2 + \mu m c^2 W'' \right] .  
\label{quart-wpp}  \end{align}

At leading order, we thus have 
\begin{align}
W = \frac{\mu m c^2 q'' }{\rho} = \frac{ (1-mc^2) q'' }{\rho} , 
\label{app-tw-a-w}  \end{align}
which provides an equation for the expected speed of the wave, 
$c=c_0$.   We consider waves which travel at speeds close to 
$c_0$, and hence, we write 
\begin{align}
c = c_0 (1+c_1 h^2) ,& \quad c_0=\frac{1}{\sqrt{m(1+\mu)}} . 
\label{app-tw-co}  \end{align}

 Adding (\ref{quart-upp}) and (\ref{quart-wpp}) together with $c$  being 
given by (\ref{app-tw-co}) and $W$ by (\ref{app-tw-a-w}) implies
\begin{align}
2 c_1 q'' &=\; \gamma q'''' + 3 b q'^2 q''  , \\ 
\gamma & = \;   \rec{12} + \frac{\mu^2}{\rho(1\!+\!\mu)^2} . 
\label{app-tw-gam-def}
\end{align}
After integrating with constants of integration set to zero, 
so that $q',q''\rightarrow0$ as $z\rightarrow\pm\infty$, 
we obtain the solution 
\begin{align}
q =\pm \sqrt{\frac{2\gamma}{b}} \tan^{-1} \left( \sinh \left( 
(y-c\tau) \sqrt{\frac{2 c_1}{\gamma}} \right) \right) . 
\end{align}
We note that both this leading order solution, and that for 
$Q\sim q+\mathcal{O}(h^2)$, are independent of $\alpha,\beta$, 
depending only on $\rho,\mu,b,c_1$, with $c$ being given by 
(\ref{app-tw-co}). 

\subsection{Cubic potential, $a\neq0$, $b=0$}
\label{tw-pure-cub-sec}

We again leave $\alpha,\beta$ arbitrary, giving the 
governing ODEs 
\begin{align}
\!\!m \frac{\dd^2 q_n}{\dd t^2} \!=& q_{n+1} \!-\! 2 q_n \!+\! q_{n-1} 
\! +\!a (q_{n+1}\!-\!q_n)^2 \!-\! a (q_n\!-\!q_{n-1})^2 
\nn \\ & - \rho (q_n-Q_n) - \alpha (q_n-Q_n)^2 - \beta (q_n-Q_n)^3 , 
\nn \\ & \\ 
\!\!\!M \frac{\dd^2 Q_n}{\dd t^2} =&\; \rho(q_n-Q_n) 
+ \alpha (q_n-Q_n)^2 + \beta (q_n-Q_n)^2 . \nn \\ & 
\end{align} 
We follow the same procedure as in \S\ref{tw-pure-quart-sec},  
namely applying (\ref{app-tw-a-s1})--(\ref{app-tw-a-s2}), 
which leads to the hence the approximating PDEs 
\begin{align}
m q_{\tau\tau} = & \; q_{yy} + \rec{12} h^2 q_{yyyy} 
+ a h (q_y^2)_y - \rho W - \alpha h^2 W^2 , \\ 
\mu m q_{\tau\tau} = & \; \rho W + \alpha h^2 W^2 + h^2 \mu m W_{\tau\tau} .  
\end{align}
Here, as well as the leading-order terms, we have retained terms 
of $\mathcal{O}(h)$ and $\mathcal{O}(h^2)$ but neglected terms 
higher than these.  We now assume a TW, writing 
\begin{align}
q(y,\tau) = h u(z), \quad W(y,\tau) = h w(z) , \quad 
z = y - c \tau , \nn \\ & 
\end{align}
to obtain 
\begin{align}
(m c^2-1) u'' + \rho w = & \; h^2 \left[ \rec{12} u'''' + a  (u'^2)' \right]  , 
\label{quad-upp} \\ 
\mu m c^2 u'' - \rho w  = & \; h^2 \mu m c^2 w'' . \label{quad-wpp}
\end{align}

At leading order, we thus have 
\begin{align}
w =\frac{\mu m c^2 u'' }{\rho} = \frac{(1-mc^2) u''}{\rho} , 
\label{app-tw-b-w}
\end{align}
which gives the same expression for $c_0$ as previously (\ref{app-tw-co}). 
Combining (\ref{quad-upp})--(\ref{app-tw-b-w}) with (\ref{app-tw-co})
\begin{align}
2 c_1 u'' & = \; \gamma u'''' + 2 a u' u'' , 
\end{align}
with $\gamma$ as defined by (\ref{app-tw-gam-def}). 
After integrating with all constants of integration set to 
zero, so that $u',u''\rightarrow0$ as $z\rightarrow\pm\infty$, 
we find the solution 
\begin{align}
u & = \; \frac{3}{a} \sqrt{ 2 \gamma c_1} \, \tanh \left( 
(y-c\tau) \sqrt{\frac{c_1}{\gamma}} \right) .  
\end{align}
As with the quartic potential case, at leading order, the solution 
has no dependence on $\alpha,\beta$, it only depends on 
$\mu,\rho,c_1,a$. 


\section{Special solutions of the Ginzburg-Landau equation (\ref{gnls})}

Due to the fully quadratic case ($\alpha \neq 0 \neq a$, considered in 
Sec.~ \ref{fully-quad-sec}) being governed by the Ginzburg-Landau 
equation rather than than the NLS (Sec \ref{a0a0-subsec}), we expect 
that the form and stability of waves in the case $\alpha \neq 0 \neq a$ 
could differ from those cases with the more standard reduction to NLS, 
where, for some lattice systems, large-time stability results have been 
derived, see for example, \cite{pego,dp,ober,HamLatDynMinE}. 

The form of some special solutions of the Ginzburg-Landau equation 
(\ref{gnls}) can be obtained from the ansatz
\beq
F_1(z,T) = A(Z) \exp \left( i \Phi(Z) + i \tilde\omega T \right) , 
\quad Z = z-uT , 
\eeq
which leads to the coupled ODEs for the real functions $A(Z),\phi(Z)$ 
\begin{align}
\Omega u A \Phi' - \tilde\omega \Omega A = &\; \eta A^3 + D_3 A'' 
- D_3 A (\Phi')^2 , \label{GLre}  \\
- \Omega u A' = & \; \zeta A^3 + D_3 A \Phi'' + 2 D_3 A' \Phi' . 
\label{GLim}
\end{align}
We seek solutions in which $A(Z)$  is single-humped and decays to 
zero in both limits $Z \rightarrow\pm\infty$.  

We introduce $S(Z)$ defined by $S'(Z) = A(Z)^4$ whereupon 
the latter equation (\ref{GLim}) implies 
\begin{align}
\frac{\dd}{\dd Z} \left( D_3 A^2 \Phi' + \half u \Omega A^2 \right) 
= - \zeta A^4 = -\zeta \frac{\dd S}{\dd Z} , \label{GLim2}
\end{align}
and the former equation (\ref{GLre}) gives 
\begin{align}
\left(\! D_3 A^2 \Phi' \!+\!  \frac{u \Omega A^2}{2} \!\right)^{\!\!2} \!= & \; 
D_3 A^4 \left[ \frac{D_3 A''}{A} \!+\! \eta A^2 \!+\! \Omega \tilde{\omega} 
\!+\! \frac{u^2 \Omega^2}{4D_3} \right] . 
\nn \\ & \label{GLre2}
\end{align}
Integrating (\ref{GLim2}) and substituting in (\ref{GLre2}), we find 
\begin{align}
\zeta^2 S^2 = &\; \frac{D_3 S'}{4}  \left[ 
\frac{D_3 S'''}{S'} - \frac{3 D_3 S''^2}{4 S'^2} + 4 \eta \sqrt{S'} 
\right. \nn \\ & \left. \qquad \qquad 
+ 4 \Omega \tilde{\omega} + \frac{u^2 \Omega^2 }{D_3} \right] . 
\label{GLreduc4}
\end{align}

Since we seek solutions in which $A(Z)$ has the form of a pulse,
$S(Z)$ will have the form of a kink-wave, which can be assumed to 
be monotonically increasing.  As (\ref{GLreduc4}) is autonomous, 
we rewrite it as a non-autonomous second-order equation using 
\begin{align}
P(S)  = & \; \frac{\dd S(Z)}{\dd Z} , \quad Z = \int \frac{\dd S}{P(S)} , 
\label{PS}
\end{align}
so that 
\begin{align}
S''(Z) = P(S)P'(S), & \quad  S'''=P(PP')'=P^2P''+PP'^2 . 
\nn \\ & 
\end{align}
We expect $P(S)>0$ with $P(0)=0$ and $P(S_1)=0$ for some $S_1>0$. 
Hence 
\begin{align}
\frac{4\zeta^2 S^2}{D_3 P } = &  D_3 ( P P'' \!+\! \rec{4} P'^2 ) + 
4 \eta \sqrt{P}  + 4 \Omega \tilde{\omega} + \frac{u^2 \Omega^2}{D_3} .  
\nn \\ & \label{GLreduc5} \end{align}
which can be  simplified by the substitution $P = R^{4/5}$
\begin{align}
\rec{5} D_3 R'' + \eta R^{-1/5} + \left( \tilde{\omega} + \frac{u^2\Omega}{4D_3} 
\right) R^{-3/5} = &\; \frac{\zeta^2 S^2 }{D_3 R^{7/5}}. 
\nn \\ & \label{GLreduc9} \end{align}

By considering the asymptotic behaviour of this 
in the limit $Z\rightarrow0$ where $A\rightarrow0$, we note 
qualitative differences between the classic NLS limit when $\zeta=0$, 
and the GL limit when $\zeta\neq0$. 
This limit corresponds to $S\rightarrow S_0>0$ and 
$P,R \rightarrow0$. 

When $\zeta=0$ the leading order balance is given by one of 
\begin{align}
\rec{5} D_3 R'' \sim & \; B R^{-3/5} , \quad 
B:= \tilde{\omega}\Omega + \frac{u^2\Omega^2}{4D_3} , \nn \\ 
\rec{5} D_3 R'' \sim & \; - \eta R^{-1/5} , 
\end{align}
the second case occurring if $B=0$. 
Writing $R(S) \sim R_0 (S_0-S)^\gamma$, 
these correspond to the equations 
\begin{align}
\gamma=\mfrac{5}{4} , \quad & Z \sim \log | S_0 - S | , 
\\ 
\gamma=\mfrac{5}{3} , \quad & Z \sim (S_0-S)^{-1/3} . 
\end{align}
Here we have made use of $R = P^{5/4}$ and (\ref{PS}), 
and find $Z\rightarrow \infty$ as $S\rightarrow S_0$, 
and hence $A \rightarrow0$.  The first case has exponential 
convergence ($A\sim \ee^{-\lambda Z}$), whilst the latter, 
algebraic decay ($A \sim z^{-4}$).

However, in the case $\zeta\neq0$, the leading order 
balance in (\ref{GLreduc9}) is 
\beq
\rec{5} D_3 R'' \sim \frac{ \zeta^2 S_0^2 }{D_3 R^{7/5}}, 
\eeq
hence we have $\gamma=5/6$ and 
$P^{5/4} = R \sim (S_0 - S)^{5/6}$, 
\beq
P = \frac{\dd S}{\dd Z}  \sim (S_0 - S)^{2/3} , 
\eeq
which implies $- Z \sim (S_0 - S)^{1/3}$, 
and $Z\rightarrow0$ as $S \rightarrow S_0$, 
rather than $Z\rightarrow \infty$ as $S \rightarrow S_0$. 
Thus, whilst the GL reduction may give rise to periodic 
waves, and waves of a more complicated form than those 
derived from the NLS, we do not expect to see single-humped 
pulse solitons, with exponentially-decaying tails.

\bibliographystyle{unsrt}


\end{document}